\documentclass[preprint]{aastex631}
\pdfoutput=1

\usepackage{amsmath}
\usepackage[caption=false]{subfig}

\newcommand{\alfven}{Alfv\'{e}n\ }
\newcommand{\elsasser}{Els{\"a}sser\ }

\newcommand{\whaivfont}{
}
\DeclareTextFontCommand{\whaiv}{\whaivfont}

\makeatletter
\providecommand\add@text{}
\newcommand\tagaddtext[1]{%
  \gdef\add@text{#1\gdef\add@text{}}}%
\renewcommand\tagform@[1]{%
  \maketag@@@{\llap{\add@text\quad}(\ignorespaces#1\unskip\@@italiccorr)}%
}
\makeatother

\shortauthors{Ashfield \& Longcope}

\graphicspath{{./}{figures/}}

\begin{document}

\title{
A Model for Gradual Phase Heating Driven by MHD Turbulence in Solar Flares
}

\accepted{January 8th, 2023}
\submitjournal{ApJ}

\author{William Ashfield IV}
\affiliation{Bay Area Environmental Research Institute, NASA Research Park, Moffett Field,
CA 94035, USA}
\affiliation{Lockheed Martin Solar \& Astrophysics Laboratory, Organization A021S, Building 252, 3251 Hanover Street, Palo Alto, CA 94304, USA}

\author{Dana Longcope}
\affiliation{Dept.\ of Physics, Montana State University,
Bozeman, MT 59717}


\begin{abstract}

\whaiv{Coronal flare emission is commonly observed to decay on timescales longer than those predicted by impulsively-driven, one-dimensional flare loop models. }
This discrepancy is most apparent during the gradual phase, where emission from these models decays over minutes, in contrast to the hour or more often observed. Magnetic reconnection is invoked as the energy source of a flare, but should deposit energy into a given loop within a matter of seconds.  Models which supplement this impulsive energization with a long, persistent {\em ad hoc} heating have successfully reproduced long-duration emission, but without providing a clear physical justification.  Here we propose a model for extended flare heating by the slow dissipation of turbulent Alfv\'{e}n\ waves initiated during the retraction of newly-reconnected flux tubes through a current sheet. Using one-dimensional simulations, we track the production and evolution of MHD wave turbulence trapped by reflection from high-density gradients in the transition region. Turbulent energy dissipates through non-linear interaction between counter-propagating waves, modeled here using a phenomenological one-point closure model. AIA EUV light curves synthesized from the simulation were able to reproduce emission decay on the order of tens of minutes. We find this simple model offers a possible mechanism for generating the extended heating demanded by observed coronal flare emissions self-consistently from reconnection-powered flare energy release.

\end{abstract}

\section{Introduction} \label{sec:intro}

Solar flares are typically characterized by the increase in radiation observed during their onset. Accepted as the result of energy released through magnetic reconnection, the sudden rise in emission across different wavelengths is considered to be an indication of impulsive plasma heating. Following the conclusion that flare loops convert magnetic energy on the order of the Alfv\'{e}nic transit time across a given loop  \citep{kopp1976,priest2002}, numerical flare loop models aiming to study the impulsive phase have commonly used energy source terms that reflect these short timescales. Many of these models have also been constrained by observations to infer energy deposition rates --- either through hard X-rays \citep{kowalski2017,graham2020} or UV emission \citep{longcope2010, qiu2012,ashfield2022} --- where the heating duration was found to last up to tens of seconds.

Although impulsively-driven models have been a successful tool for explaining a number of flare phenomena, their failure to reproduce gradual phase emission is widely recognized. Observations in soft X-rays and EUV have shown the gradual decay of hot coronal plasma to last between tens of minutes to several hours following the rise phase. The cooling of coronal plasma, marked by the successive peaks of emission light curves from increasingly cooler ion species, happens through a combination of heat conduction  \citep{culhane1970,culhane1994} and radiative losses \citep{aschwanden2001,vrsnak2006}. Compared to the characteristic cooling timescales modeled by individual flare loops \citep[e.g. minutes,][]{kerr2020}, the sustained duration of observed coronal emissions is considerably longer than it would be if regulated by cooling alone.

The discrepancy between modeled and observed decay rates during the gradual phase has led many to believe that additional heating is required beyond the impulsive phase to offset the effects of radiative losses \citep{withbroe1978,svestka1989,ryan2013,sun2013}. A popular interpretation of this heating has been the successive energization of multiple individual flux tubes within a flare arcade ---  so-called multi-loop models \citep{dere1979,hori1997,warren2006,reep2022} --- with each loop emulating the prescribed impulsive energy release. While these models are consistent with observation, several investigations were unable to rectify the lack of sustained coronal emissions \citep{reeves2002,qiu2012,liu2013,kerr2020}. 

In response to \whaiv{this} persistent discrepancy, \cite{qiu2016} recently developed a model where individual loops were driven with a two-part heating profile  consisting of an impulsive energy release followed by a prolonged, low-rate heating. This \textsl{slow-tail} heating, lasting on the order of 20\,minutes, was able to successfully forward-model key characteristics of EUV lightcurves measured using the \textsl{Atmospheric Imaging Assembly} \citep[AIA;][]{lemen2012}, including long cooling timescales. While motivated by flare UV footpoint emissions that display similar, two-phase behavior \citep[see also][]{cheng2012,qiu2013,zhu2018}, the heating profiles introduced in this work were only \textsl{ad hoc}. Physical justifications behind extended loop heating were not provided outside of a few, potentially viable scenarios.

Several mechanisms for producing such extended heating rates alluded to in \cite{qiu2016} and other investigations are contingent upon post-reconnection flare loops retracting under magnetic tension and releasing energy \citep{forbes1996,linton2006}. One such speculation is the continual heating from slow-shocks generated by the resistance loops receive from earlier formed loops that exist lower in the flare arcade \citep{cargill1982,cargill1983}. Another related mechanism is the excitation of MHD waves by reconnection outflows \citep{aschwanden2006} that then decay and heat plasma within a post-reconnection loop over timescales comparable to slow-tail heating \citep{wang2011}. 

Central to both of the scenarios described above is the interaction between retracting flare loops and their surroundings. Previous investigations of outflows driven by magnetic tension have modeled flux tubes moving through an ideal current sheet unobstructed, and have therefore retracted at the local \alfven speed \citep{forbes1983,longcope2015}. Observed outflows --- inferred from the motion of supra-arcade downflows \citep[SADs;][]{mckenzie1999,savage2011,reeves2017} interpreted to be the wakes behind retracting flux tubes \citep{savage2012} --- however, move at fractions of the Alfv\'{e}n speed. In the case of a non-ideal current sheet, such as that with a high-$\beta$ plasma \citep{scott2016}, a retracting loop would likely undergo a loss of momentum and speed as it imparted work on the surrounding medium. This interaction, modeled as a simple aerodynamic drag force, was recently found to slow outflow speeds to within their observed ranges \citep{unverferth2021}.

The interpretation of SADs as the wakes of retracting flux tubes experiencing a drag force would also suggest the generation of turbulence, given the aerodynamic analogy. 
In fact, evidence for turbulence has been reported in current sheets containing SADs, inferred through measurements of non-thermal line broadening \citep{ciaravella2008,warren2018,cheng2018} and local correlation tracking \citep{mckenzie2013}. Separate observations using AIA and the Extreme-ultraviolet Imaging Spectrometer \citep[EIS;][]{culhane2007} have also shown turbulence to coexist in regions with hot, $\sim30$\,MK downward moving loops, but did not make reference to SADs \citep{imada2013,doschek2014}. Using observations of several strong flares, 
\cite{larosa1993} found that the interactions between individual flux tubes and interactions with a diffusive current sheet are likely to develop outflows with a turbulent structure. Moreover, they also found MHD turbulence to be capable of rapidly thermalizing bulk kinetic energy via a turbulent cascade, thus becoming a mechanism for impulsive phase energy release. It was further suggested by \cite{jiang2006} that plasma wave turbulence could produce heating into the decay phase, indicated by the long cooling times of loop-top HXR sources. Although the interconnected behavior of retracting flux loops, turbulence, and extended flare heating has been well established by observation, it has yet to be modeled in detail.

The present work aims to develop a self-consistent mechanism for gradual-phase energy release. Motivated by the observations of SADs, we create a relatively simple model for the excitation of MHD turbulence along a \textsl{thin flux tube} (TFT) as it retracts through a current sheet. The retracting tube experiences a drag force, which converts energy from the tube into an internal population of turbulent \alfven waves that then propagate along the tube and reflect off high-density gradients in the transition region. Turbulent energy is dissipated through the non-linear interaction between counter-propagating waves, thus creating a local heat source within the tube. A model for drag with turbulent transport is outlined in Section \ref{sec:cond}. The behavior of our model is then demonstrated using an initial reference simulation in Section \ref{sec:ref}. Using the simulation results, synthetic AIA EUV emissions are created by which our model can be qualitatively compared to observation. The parameters critical to the duration of turbulent energy release --- the fraction of drag power converted to turbulent \alfven waves and the energy correlation length of the propagating waves --- are explored in Section \ref{sec:ext}. A final simulation run with a set of optimized parameters was found to extend the duration of EUV emission past 40\,minutes. The duration of the heating produced from turbulent dissipation was also found to persist long after the retraction had ended. The results presented in this paper point to a physical mechanism that can produce late-phase heating required to sustain coronal flare emission for the first time.

\section{Flare Model with Alfv\'en wave Turbulence} \label{sec:cond}

\begin{figure*}
    		\centering
    	\subfloat[]{
    		\includegraphics[width=0.47\textwidth]{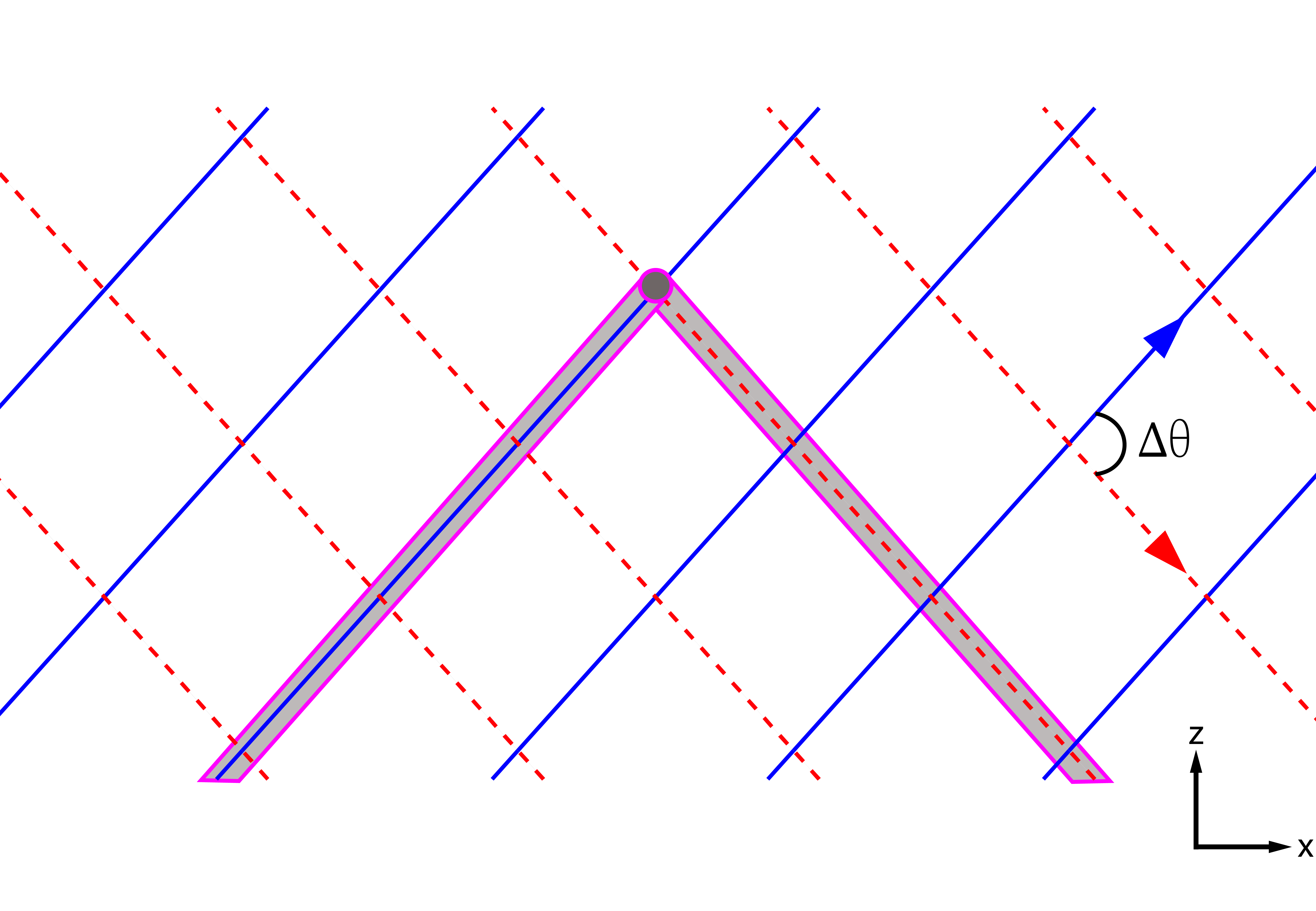}
    		\label{loop2}
    	}
            \centering
    	\subfloat[]{
    		\includegraphics[width=0.47\textwidth]{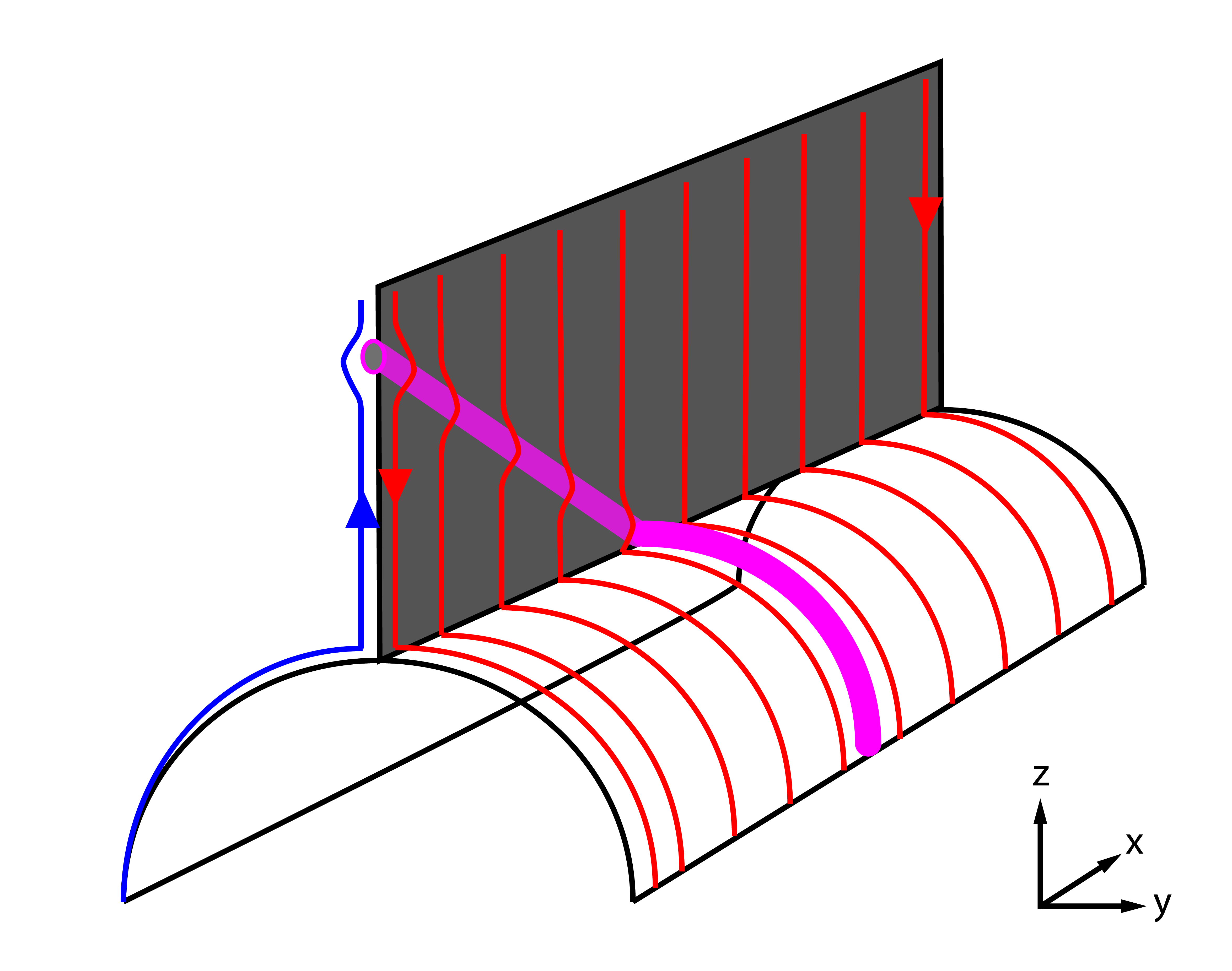}
    		\label{loop3}
    	}
    \caption{Schematic of a reconnected flux tube embedded in a current sheet. (a) Face-on view \whaiv{of the current sheet. External magnetic fields on either side of the current sheet --- the blue and red-dashed lines --- are skewed according to the sheer angle $\Delta\theta$. The grey tube shows the geometry of the newly formed flare loop immediately following reconnection.  (b) Perspective orientation illustrating the deflection of the external magnetic fields around the flux tube, with the tube sliced through its apex. The grey area corresponds to the current sheet shown in (a). Only the $\hat{\textbf{z}}$-component of the external magnetic fields is shown in the latter. }}
    \label{fig:schem}
    \end{figure*}

In order to model the energy released by reconnection and its conversion to MHD turbulence, we employ the
TFT equations described by \cite{longcope2009}. The model assumes localized reconnection has already occurred within a small diffusion region, within the current sheet, to create a closed magnetic flux tube. The moment following reconnection is illustrated in Figure \ref{fig:schem}. Two layers of equilibrium magnetic field are separated by a current sheet and have directions differing by a shear angle $\Delta\theta$. The newly reconnected loop, shown in grey, is embedded within the current sheet that lies in the $x$--$z$ plane. It is at this point the TFT model begins. Mechanisms of the reconnection process itself are not considered here. 
    
Once the bent tube is initialized, it retracts through the current sheet under a magnetic tension force. The internal magnetic pressure of the tube is balanced by \whaiv{the} magnetic pressure of the external magnetic fields separated by the current sheet,  setting the magnetic field strength inside the tube. In the present case, we assume the external magnetic fields to be uniform, such that the magnetic field $B$ is also uniform along the tube.

The TFT equations govern the evolution of the axis of a flux tube, $\textbf{r}(\ell,t)$, where $\ell$ is the parameterized length coordinate of the curved tube. The fluid velocity of the plasma confined within the tube, $\textbf{v}=d\textbf{r}/dt$, is advanced according to the momentum equation \citep{longcope2015} 
\begin{equation}
\label{eq:momen}
    \rho\frac{d\textbf{v}}{dt} = \Bigg( \frac{B^2}{4\pi} - p \Bigg) \frac{\partial\hat{\textbf{l}}}{\partial \ell} - \hat{\textbf{l}}\frac{\partial p}{\partial \ell} + \rho \textbf{g} +  \tfrac{4}{3}\mu\frac{\partial}{\partial \ell}  \Bigg(\hat{\textbf{l}}\hat{\textbf{l}} \cdot \frac{\partial \textbf{v}}{\partial \ell} \Bigg),
\end{equation}
where $p = \rho k_BT/\bar{m}$ is the plasma pressure and $\bar{m} = 0.593\ m_p$ is the mean particle mass of the plasma, assumed to be fully ionized with a coronal abundance. For a fluid element along the tube with differential length $d \ell$ and mass per unit flux $dm$, the mass density is given by $\rho = B (dm/\,d\ell)$.

Outside of gravitational force $\rho \textbf{g}$, which acts in the downward direction, forces acting on a fluid element in Equation (\ref{eq:momen}) are dictated by the tangent vector, $\hat{\textbf{l}} = \partial \textbf{r}/\partial \ell$. The first term on the rhs describes the magnetic tension force along the tube’s curvature vector, $\partial \hat{\textbf{l}} / \partial \ell$, perpendicular to the tube's axis. The second is the gas pressure directed along the tube. Viscous interactions between fluid elements are described by the final term, with $\mu = 0.012\ \kappa_{\text{sp}}(T) / c_v$ being the temperature-dependent dynamical viscosity coefficient and the classical Spitzer-H{\"a}rm thermal conductivity being $\kappa_{\text{sp}} = 10^{-6}\ T^{5/2}$, in conventional cgs units.

The energy of the tube is advanced according to the temperature of a fluid element
\begin{equation}
\label{eq:en}
    c_v \rho \frac{dT}{dt} = -p\Bigg( \hat{\textbf{l}} \cdot \frac{\partial \textbf{v}}{\partial \ell} \Bigg) + \tfrac{4}{3}\mu \Bigg( \hat{\textbf{l}} \cdot \frac{\partial \textbf{v}}{\partial \ell} \Bigg)^2  - n_e^2 \Lambda(T)  + \frac{\partial}{\partial\ell}\left(\kappa \frac{\partial T}{\partial \ell}\right), 
\end{equation}
where $c_v = 3k_B/2\bar{m}$ is the specific heat and $n_e = 0.874\ (\rho/m_p)$ is the electron number density. The first term on the rhs constitutes the adiabatic compression done on the gas. This term is followed by viscous heating, which arises from the loss of kinetic energy in Equation (\ref{eq:momen}). Optically thin radiative losses are given by the third term, where the radiative loss function $\Lambda (T)$ is taken from the output of CHIANTI 7.1 \citep{landi2012}. The final term describes the field-aligned thermal conduction. A flux limiter is included in the model, such that the thermal conductivity is restricted to the theoretical electron free-streaming limit at sufficiently large temperature gradients \citep{longcope2015}.

\subsection{Drag Force and Alfv\'{e}n Wave Turbulence} 

The TFT model describes the evolution of flare loops retracting under a tension force brought about by a change in magnetic topology. Previous studies assumed the retraction occurs without any interaction with the background current sheet, and therefore saw typical outflow velocities reaching the local \alfven speed on the order of 3\,Mm\,s$^{-1}$ \citep{longcope2018, unverferth2020}.  Alfv\'enic outflow speeds are common to all idealized models of magnetic reconnection, but are not supported by much observational evidence.  Supra-arcade downflows \citep[SADs][]{mckenzie1999} are often taken to be signatures of post-reconnection retraction, but always seem to move well below the local Alfv\'en speed \citep{savage2011,savage2012}.

Based on observations of sub-Alfv\'enic outflow speeds it has been proposed that the retracting flux must somehow interact with the surrounding plasma, possibly by deforming the flux outside the sheet or entraining plasma \citep[see Figure \ref{loop3} and discussions in][]{linton2006,scott2013}. Most forms of interaction would leave the surrounding plasma with greater magnetic or kinetic energy, at the expense of the retracting tube.  This would appear as some kind of drag force on the retracting flux, removing some \whaiv{of} its energy and reducing the retraction speed.  \cite{unverferth2021} investigated this possibility, introducing a drag force modeled \whaiv{using a} high Reynolds number aerodynamic drag \citep{choudhuri1987}
\begin{equation}
\label{eq:f_d}
    \textbf{f}_d = -D\, \rho\ |\textbf{v}_\perp|\ \textbf{v}_\perp.
\end{equation}
 Here $\textbf{v}_\perp = \textbf{v}-\hat{\textbf{l}}(\hat{\textbf{l}} \cdot \textbf{v})$ is the component of fluid velocity perpendicular to the flux tube and $D$ is a constant proportional to the traditional drag coefficient. Because the interaction between the flux tube and the current sheet is likely more intricate than the conventional drag force exerted by a neutral fluid on a rigid body, $D$ is taken to be a free parameter to capture these unknown complexities \citep{unverferth2020}.  This force, per unit area, is added to the momentum equation (\ref{eq:momen}).

The force exerted by the surrounding plasma ultimately opposes the acceleration of the loop by transferring momentum and energy away. Energy is therefore removed from the retracting loop at a rate of 
\begin{equation}
\label{eq:pow}
    P_d = \rho^{-1}\, \mathbf{v} \cdot \textbf{f}_d \leq 0,
\end{equation}
which is always negative.  The energy is not really lost, but must appear in some form in the surrounding plasma.  We assume here that it takes the form of MHD turbulence, of which some fraction remains on the tube, but at length scales heretofore unresolved.  We therefore assume a fraction of $P_d$ takes the form of unresolved \alfven wave turbulence occurring on the tube itself, even as it retracts.

\subsection{Evolution of Alfv\'{e}n Wave Turbulence}

Gaining inspiration from turbulent transport models of the solar wind \citep[i.e.][]{marsch1989,matthaeus1999,dmitruk2001,verdini2007,lionello2014}, the equations used for the evolution of turbulent wave energy in this work are developed by first decomposing the velocity and magnetic fields into large-scale contributions, denoted by capitals, and small-scale perturbations, denoted by lower case:
\begin{gather}
    \textbf{v} = \textbf{U} + \textbf{u}\\
    \textbf{B} = \textbf{B}_0 + \textbf{b} = B\hat{\textbf{l}} + \textbf{b}  .
\end{gather}
The perturbations can then \whaiv{be} collected in terms of the \elsasser variables,
$\textbf{z}_\pm  \equiv \textbf{u} \pm \textbf{b}/\sqrt{4 \pi \rho}$. 

The evolution of the fluctuations is expressed by the scale-separated, linearized MHD equations in their \elsasser representation \citep{zhou1990_a,zhou1990_c,zank2012}
\begin{equation}
\label{eq:zpm}
     \frac{D \textbf{z}_\pm}{D t} = \pm(\textbf{V}_A \cdot \nabla)\textbf{z}_\pm + \tfrac{1}{2}(\textbf{z}_\mp - \textbf{z}_\pm)  \nabla \cdot \bigg( \frac{\textbf{U}}{2} \pm \textbf{V}_A   \bigg).
\end{equation}
Here, $\textbf{V}_A = \textbf{B}_0/\sqrt{4\pi\rho}$ is the local \alfven speed along the loop. We assume the large-scale fields of \textbf{U} and \textbf{B} vary slowly across the flux tube and have therefore ignored the gradient terms in Equation (\ref{eq:zpm}). The source and sink terms for the \elsasser variables are also temporarily neglected here and are instead discussed in detail below.

The aggregate energy densities of the turbulent \alfven waves are then found by spatially averaging over the small-scale fluctuations. The energy per unit mass of the two turbulent waves is given by:
\begin{equation}
    w_\pm = \tfrac{1}{4}\langle \textbf{z}_\pm \cdot \textbf{z}_\pm \rangle
\end{equation}
with the total energy density of the perturbations being $ w_{\mathrm{tot}} = w_+ + w_- = \langle u^2/2 \rangle + \langle b^2/8\pi\rho \rangle$. By averaging over the perturbations, we remove their explicit dependence from the system and instead model their collective evolution implicitly.

Taking the inner product of $\textbf{z}_\pm$ with Equation (\ref{eq:zpm}) for $\textbf{z}_\pm$, respectively, gives the expression for each \elsasser energy $w_\pm$:
\begin{equation}
\label{eq:lag_w}
    \frac{D w_\pm}{D t } = \pm (\textbf{V}_A \cdot \nabla)w_\pm - w_\pm \nabla \cdot \bigg( \frac{\textbf{U}}{2} \pm \textbf{V}_A  \bigg) + R_\pm.
\end{equation}
Here, $R_\pm$ is a term  proportional to $\langle \textbf{z}_+ \cdot \textbf{z}_- \rangle$ that accounts for wave reflection. Analogous to the `mixing' effects described in \cite{zhou1990_c}, the interaction between opposite \elsasser variables allows for energy to be redistributed between the two populations, such that $w_\pm$ will generate counter-propagating $w_\mp$. This process is linked to large-scale gradients in the system, and is likely to be most effective where $\partial \mathrm{ln} \rho / \partial \ell$ is large \citep{holloweg1981,velli1993}. 
Computing $R_{\pm}$ self-consistently would require additional equations beginning with one for the evolution of $\langle \textbf{z}_+ \cdot \textbf{z}_- \rangle$.  In the interest of simplicity, we forego this approach and instead set $R_\pm=0$ and account for wave reflection with a boundary condition described below. 

Following the TFT model, we parameterize the wave energy equations according to length coordinate $\ell$. As the large-scale fields are taken to vary only in the direction parallel to $\hat{\textbf{l}}$,  divergences in $\textbf{U}$ and $\textbf{V}_A$ reduce to spatial derivatives in $\ell$. The full turbulent transport equations in their conservative forms are then
\begin{equation}
\label{eq:w}
    \frac{dw_\pm}{dt}  = \pm v_A \frac{\partial w_\pm}{\partial \ell} -\tfrac{1}{2} w_\pm \frac{\partial v_\parallel}{\partial \ell} \mp w_\pm \frac{\partial v_A}{\partial \ell} + S_\pm + NL_\pm , 
\end{equation}
for $v_\parallel = \hat{\textbf{l}} \cdot \textbf{v}$. Because the terms in the above expression are dependent only on system variables aligned with the mean magnetic field, Equation (\ref{eq:w}) describes the energies of small-scale \alfven waves propagating along the field in the $\pm \hat{\textbf{l}}$ direction. Propagation of the wave energies is described by the first three terms on the rhs. The first is the advective term, showing how the wave energies propagate at the \alfven speed. The second and third terms describe work done on the waves by compression of the plasma and magnetic pressures against that of the wave, respectively.

\subsection{Sources and Sinks of Alfv\'{e}n Wave Turbulence}

The last two terms of Equation (\ref{eq:w}), absent from Equation (\ref{eq:lag_w}), are added to describe the source, $S_\pm$, and sink, $NL_\pm$, of the turbulent energies, respectively. The source, as described above, is a fraction, $f_{\text{turb}}$, of the power the tube loses through drag as it retracts through the current sheet described by Equation (\ref{eq:pow}):
\begin{equation}
\label{eq:src}
    S_\pm = -\tfrac{1}{2}\,f_\mathrm{turb}\,P_d.
\end{equation}
Here the $\tfrac{1}{2}$ prefactor assumes the input energy is divided equally between the counter-propagating wave species.

The loss of turbulent energy in the system is modeled according to the simple phenomenological decay rate used in numerous MHD turbulence investigations \citep[e.g.][]{hossain1995,matthaeus1999,dmitruk2001,zank2012}. Following one-point closure models for hydrodynamic turbulence first described by \cite{karman1938}, the energy loss arises from the non-linear interaction between the counter-propagating waves. This interaction is assumed to produce a cascading spectrum of turbulent \alfven waves --- defined by a set of wave numbers that correspond to the cascade of energy-containing eddies into increasingly smaller scales --- that ultimately results in the dissipation of turbulent energy. 

The dissipation rate for the \elsasser energies is expressed as
\begin{equation}
\label{eq:nl}
    NL_\pm = - \frac{w_\pm \sqrt{w_\mp}}{\lambda_\perp},
\end{equation}
where ${\lambda_\perp}$ is the single similarity length scale that characterizes the transverse dimensions of the energy-containing eddies for both the rightward and leftward propagating waves. As such, ${\lambda_\perp}$ is the characteristic length scale that couples the non-linear spectral transfer between the counter-propagating modes, and can thus be thought to correspond to the correlation length between the two energy populations \citep{batchelor1953}. Although turbulence is typically described by a range of length scales (i.e. wave numbers), this simple one-point model assumes the decay can be represented by a single non-linear term.  

We assume that all energy lost from the turbulence is ultimately thermalized and appears as a heat source for the plasma.  This is achieved simply by adding the term 
\begin{equation}
\label{eq:hturb}
   H_\text{turb} = \rho\, \Bigl(\, NL_+ + NL_- \Bigr),
\end{equation}
to the rhs of Equation (\ref{eq:en}). With this additional term, our model establishes a connection, albeit indirect, between the energy lost to drag and a heating rate that can be used to explain long-duration EUV emission seen during the gradual phase of flares.

The TFT equations are further modified by considering the effects of turbulence on the tube's thermal conductivity. As magnetic field perturbations will lengthen the field lines along which thermal electrons carry heat, the thermal conductivity will consequently decrease. We account for this suppression by modifying the classical Spitzer-H{\"a}rm conductivity
\begin{equation}
\label{eq:sur}
    \kappa_{\text{sp}} \rightarrow \kappa_{\text{sp}}^{(\text{turb})} = \frac{\kappa_{\text{sp}}}{1+\,\rho(w_+ + w_-)/B^2},
\end{equation}
such that the turbulent energy will directly impede the heat flux in the tube. Although this is a relatively simple addition to the model, it attempts to address the notion of heat conduction suppression via turbulence in a self-consistent manner. 

Finally, we achieve turbulent wave reflection from the chromosphere using the boundary conditions
\begin{equation}
\label{eq:ref}
    w_-(\ell_0) = \eta\,w_+(\ell_0) \quad , \quad w_+(\ell_1) = \eta\, w_-(\ell_1), 
\end{equation}
where $\ell_0$ and $\ell_1$ are the left and right boundaries, respectively. Because reflections are likely to occur in the transition region where temperature and density gradients are large, these boundaries are set by a characteristic temperature $T_{\mathrm{TR}}$ such that $\ell_{0 (1)} = $ min(max)\,\{$ T > T_\mathrm{TR}$\}. In this case, waves incident on the transition region at either end of the loop will be transformed into their respective counter-propagating waves, effectively trapping the wave energy in the corona. Furthermore, the boundary condition assumes a reflection coefficient of $\eta \leq 1$ to account for energy lost from wave transmission. The value of this coefficient is taken as a free parameter and is discussed, along with the other free parameters of our model, in the following section.

\section{Simulation with Alfv\'en wave Turbulence} \label{sec:ref}

To illustrate the behavior of our model, we construct and run a single reference simulation.  Rather than attempt to model a particular observation, we aim at properties typical of a long-duration event, including peak temperatures above 20 MK and flux retraction speeds of order 500 km/s.  We attempt to extend the cooling time to values comparable to those typically found.  For concreteness, we use values reported in \cite{qiu2016}, since they also infer input energy.  To make a meaningful comparison, we choose parameters to give the simulation the same energy flux as they report.  The run and its parameters are discussed alongside their motivations below.

\subsection{Initial Conditions} \label{init}

Prior to the simulation run, a flux tube is initialized in a configuration analogous to a bent field line created from reconnection. The tube is confined to the $x$--$z$ plane, \whaiv{corresponding to} the current sheet that separates uniform layers of magnetic flux differing by shear angle $\Delta\theta = 120^\circ$, as illustrated by Figure \ref{loop2}. The initial tube is therefore bent at its apex by $180^\circ - \Delta\theta = 50^\circ$, and is set to have a uniform magnetic field of magnitude $B = 100$\,G. Because this process forms two straight segments joined at a single point, the apex of the loop is rounded to be a semi-circle composed of eight grid points to prevent issues arising from under-resolution. 

Once bent, the flux tube is initialized into three components: a coronal loop and two footpoints attached to a rudimentary chromosphere. The pre-flare chromosphere is stratified under gravity with pressure increasing exponentially according to scale height $H=500$\,km. Taken to be isothermal with temperature $T_{\mathrm{min}}$=0.01\,MK, the chromosphere primarily serves as a mass reservoir of cool, dense plasma for evaporated material. The coronal region of our flux tube is structured according to the relations given in \cite{rosner1978}. Defined by an apex temperature set to $T_{\mathrm{co},0}$=1.3\,MK, the initial corona is configured in isobaric equilibrium maintained by an \emph{ad hoc} volumetric heating input. The heating required to maintain equilibrium, \whaiv{however}, is only used during the initialization process. Subsequent evolution of the loop does not contain this heating term in order to more directly observe the consequences of turbulent heating induced along the tube. 

The initial length of the tube is calculated from the amount of flare energy we want to be released by the system. In the TFT model, the source of this energy comes from contracting magnetic field lines, releasing magnetic free energy into other forms as the flux tube retracts. The magnetic energy per unit of magnetic flux of the tube is \citep{longcope2015}
\begin{equation}
    W_M = \frac{1}{4\pi} \int B[\textbf{r}(\ell)]d\ell.
\end{equation}
Because the strength of the magnetic field is fixed, a change in magnetic energy is therefore powered by a reduction in the tube's length. A net length decrease $\Delta L$, releases flare energy, per magnetic flux, 
\begin{equation}
\label{eq:efl}
    E_{\mathrm{fl}} = \frac{B\Delta L}{4 \pi} \sin^2\bigg({\frac{\Delta \theta}{4}}\bigg).
\end{equation}
Here, the sine-squared factor represents the fraction of magnetic energy converted to kinetic energy parallel to the axis of the tube through rotational discontinuities that form during retraction \citep{longcope2015}. This mechanism constitutes the primary mode of magnetic energy conversion in previous studies using the TFT model to investigate reconnection dynamics, with and without drag \citep{guidoni2010,unverferth2021}.

To make contact with the work done in \cite{qiu2016}, we incorporate the same flare energy deposited by the impulsive part of their heating profile, $H(t)$, into the energy released in our model\footnote{The heating rate, given by Equation (6) in Section 4 of \citet{qiu2016}, is a piecewise expression composed of three Gaussians. The first two describe the impulsive flare energy release as inferred from AIA\,1600\,\AA\ ribbon emission observations, while the third is appended to model gradual-phase heating. Here we integrate the first two, along with the same parameter values described in the same section, to calculate a total energy per unit area.}. By integrating over the heating rate,  the total energy per unit area is calculated to be $\int H(t) dt = \zeta E_{\mathrm{fl}} B = \zeta 2.9\times10^{11}$\,erg\,cm$^{-2}$. The prefactor $\zeta$ is introduced to account for imperfect conversion between magnetic energy and other forms. Assuming the primary energy driving flare phenomena arises from MHD turbulence, as we argue below, the conversion is ultimately related to the fraction of power lost through drag, such that $\zeta = 1/f_\mathrm{turb}$. We initially take this fraction to be reasonably conservative at $f_\mathrm{turb}$=0.2, giving a total energy release of $1.45\times10^{12}$\,erg\,cm$^{-2}$.

Equation(\ref{eq:efl}) is then solved for $\Delta L$, giving $\Delta L$ = 73\,Mm. The loop is initialized with length $L_0 = L_\mathrm{f} + \Delta L$, with $L_\mathrm{f}$ being the final length of the loop once retraction is complete. We choose $L_\mathrm{f}$ to be 93\,Mm, which gives an initial loop length of $L_0$=166\,Mm. 

\subsection{Parameters of the Drag Model} \label{lambda}

Although the total energy released by the loop through retraction is set by Equation (\ref{eq:efl}), the inclusion of a drag force in our model changes the nature of energy release in comparison to previous TFT models. In these earlier works, parallel kinetic energy is responsible for shocks that compress and thermalize the plasma via viscous heating, while the energy found in perpendicular flows is found to have little effect on the overall magnetic energy conversion. \cite{unverferth2021} showed that kinetic energy will be effectively removed from the tube by including drag. While this reduction was shown to affect the perpendicular kinetic energy primarily, rotational discontinuities were also shown to be less effective at driving parallel flows as well, thus limiting the available energy to be converted to heat. In the case \whaiv{of} drag-induced \alfven wave turbulence, perpendicular kinetic energy is no longer lost. Instead, a fraction is reintroduced to the tube, as dictated by Equations (\ref{eq:pow}) and (\ref{eq:src}). We expect the subsequent transfer of turbulent energy to heat from dissipation will capture a portion of the magnetic energy that would otherwise have gone into perpendicular flows and be a source of heat comparable to, if not greater than, heat driven by parallel plasma flows. 

The drag coefficient, $D$, in Equation (\ref{eq:f_d}) controls both the amount of turbulent energy siphoned from perpendicular kinetic energy and the rate of retraction in the $-\hat{\textbf{z}}$ direction. As described in the introduction, drag was first motivated by observations of reconnection outflows \citep[e.g. SADs][]{savage2012}, and serves as the mechanism by which sub-Alfv\'{e}nic speeds can become consistent with Petschek-like reconnection models. Here we set $D = 50\,\mathrm{Mm}^{-1}$ in order for our model to agree with measured outflow speeds. An initial investigation using this coefficient shows the loop retracting at speeds $\sim$500\,km\,s$^{-1}$, which is comparable to the average velocity of SADs measured in \cite{longcope2018}.

\whaiv{The turbulence model includes several new parameters, used here for the first time.  These will be varied in order to understand their effects, but we first describe their significance and plausible ranges.  We will then return in the discussion to consider what their variation has revealed about the physics of turbulence.}

The turbulent energy dissipation rate is set by the energy correlation length, $\lambda_\perp$.
Investigations into the effects of turbulence on coronal heating \whaiv{have} assumed $\lambda_\perp$ to be on the order of inter-network spacing of 30\,Mm \citep{matthaeus1999,downs2016}. \cite{abramenko2013} measured the characteristic length of the energy-containing structures using local correlation tracking in the photosphere, finding $\lambda_\perp$ to lie in the range of 0.6–2\,km. Although these studies deal with the issue of coronal heating driven by convective motions in the photosphere, it is not unreasonable to take the correlation length in the coronal to be of similar values. If we instead take $\lambda_\perp$ to be on the order of the diameter of a given flux tube, then $\lambda_\perp \sim 0.1-2$\,Mm, given the distribution of loop widths measured using images taken of coronal loops with AIA and the Hi-C imager \citep{aschwanden2017}. 

There is an alternative approach through the observed decay time of Alfv\'en wave energy.  If that decay is due only to \whaiv{a} non-linear turbulent cascade, then the decay time will be given by the eddy turnover time
\begin{equation}
\label{eq:eddy}
    \tau_\mathrm{nl} = \frac{\lambda_\perp}{\langle v_\mathrm{nth} \rangle} = \frac{\lambda_\perp}{\sqrt{w}}.
\end{equation}
Investigations into the non-thermal broadening of EIS Fe\,\textsc{xxiv}\,255\,\AA\ lines during flares have seen spatially averaged non-thermal velocities $\langle v_\mathrm{nth} \rangle$ on the order of 60-100\,km\,s$^{-1}$ decay to pre-flare values in roughly 20 minutes \citep{kontar2017,stores2021}. Setting  Equation (\ref{eq:eddy}) to this time, and using this turbulent velocity, suggests a correlation length of $\lambda_\perp = 30-150$\,Mm. 

There is at least one other aspect to include in our interpretation of  $\lambda_{\perp}$.  Traditional derivations of the non-linear dissipation captured by Equation (\ref{eq:nl}) assume the counter-propagating waves are uncorrelated with one another.  Our waves, however, are trapped between reflecting ends and are likely to be correlated.  This fact could be accounted for by one more multiplicative factor, $0<\xi<1$, in Equation (\ref{eq:nl}).  Rather than including one more free parameter, we combine the two into a single free parameter $\lambda'_{\perp}=\lambda_{\perp}/\xi$, which will thus be larger than the turbulent length scale.  We hereafter interpret $\lambda_{\perp}$ as $\lambda'_{\perp}$.

Altogether, the above analysis produces a very rough range of $\lambda_\perp = 0.1$--150\,Mm. Nonetheless, given the simplicity of this model, these values are only meant to demonstrate that a variety of length scales can be used to characterize turbulent energy dissipation, while also being grounded in observation. For the purpose of a single reference model, $\lambda_\perp$ is initially set to 2\,Mm\whaiv{--- a value consistent with the diameter of observed coronal loops. This choice thereby assumes the energy correlation length to be on the order of the small-scale transverse fluctuations along the flux tube.} The effects of different length scales on the duration of turbulent heating are explored further in the following section.

Lastly, the initial value of the reflection coefficient for the propagating energy waves is set to be $\eta = 0.95$. This value, albeit close to its upper threshold of 1, is chosen so that energy losses arising from turbulent dissipation can be distinguished from losses due to energy escape through the transition region. The characteristic temperature $T_\mathrm{TR}$ defining the reflection boundaries is chosen to correspond to regions with a high-density gradient. For our tube initialized with $T_{\mathrm{co},0}$=1.3\,MK, we therefore set $T_\mathrm{TR}$= 0.5\,MK.

\subsection{The Reference Simulation} 

With the flux tube initialized as described above, Equations (\ref{eq:momen}) (\ref{eq:en}), and (\ref{eq:w}) are solved using the \textsc{PREFT} numerical code as documented in \cite{longcope2015}. Fluid elements along the tube are represented by cells on a 1D Lagrangian grid, with the density and differential length of each calculated to ensure each has a constant mass per unit flux. Aside from the evolution of temperature, which is advanced semi-implicitly, all expressions are advanced explicitly. In order to ensure stability in the system, each time step is chosen such that the Courant–Friedrichs–Lewy conditions are satisfied. Initial parameters and the resulting properties of the reference simulation are summarized in Table \ref{tbl}.

\begin{table}[t!]
\centering
\caption{\label{tbl}}
\begin{tabular}{ l c c c }
\multicolumn{4}{c}{Simulation properties of the reference and optimized runs} \\
\hline
\hline
  & & Reference & Optimized \\ 
\textsc{Drag Parameter} &  & &   \\ 
\hline
$f_\mathrm{turb}$ & ... & 0.2 & 0.6 \\
$\lambda_\perp$ & Mm  & 2 & 100 \\ 
$D$ & Mm$^{-1}$ & 50 & 50 \\ 
$\eta$ & ... & 0.95 & 0.95 \\
\textsc{Model Result} &  &  &   \\ 
\hline
$T_\mathrm{peak}$ & MK & 41  & 32 \\
Peak $\dot{Q}$  & $10^{10}$\,erg\,cm$^{-2}$\,s$^{-1}$ & 2.15 &  1.02 \\
Retraction time & min. & 7.5 & 8.1 \\
$\dot{Q}$ duration & min. &  13 & 41  \\
EUV duration & min. & 32  & 40  \\
\hline
\end{tabular}
\end{table}

\begin{figure}
\centering
\includegraphics[width=0.66\textwidth]{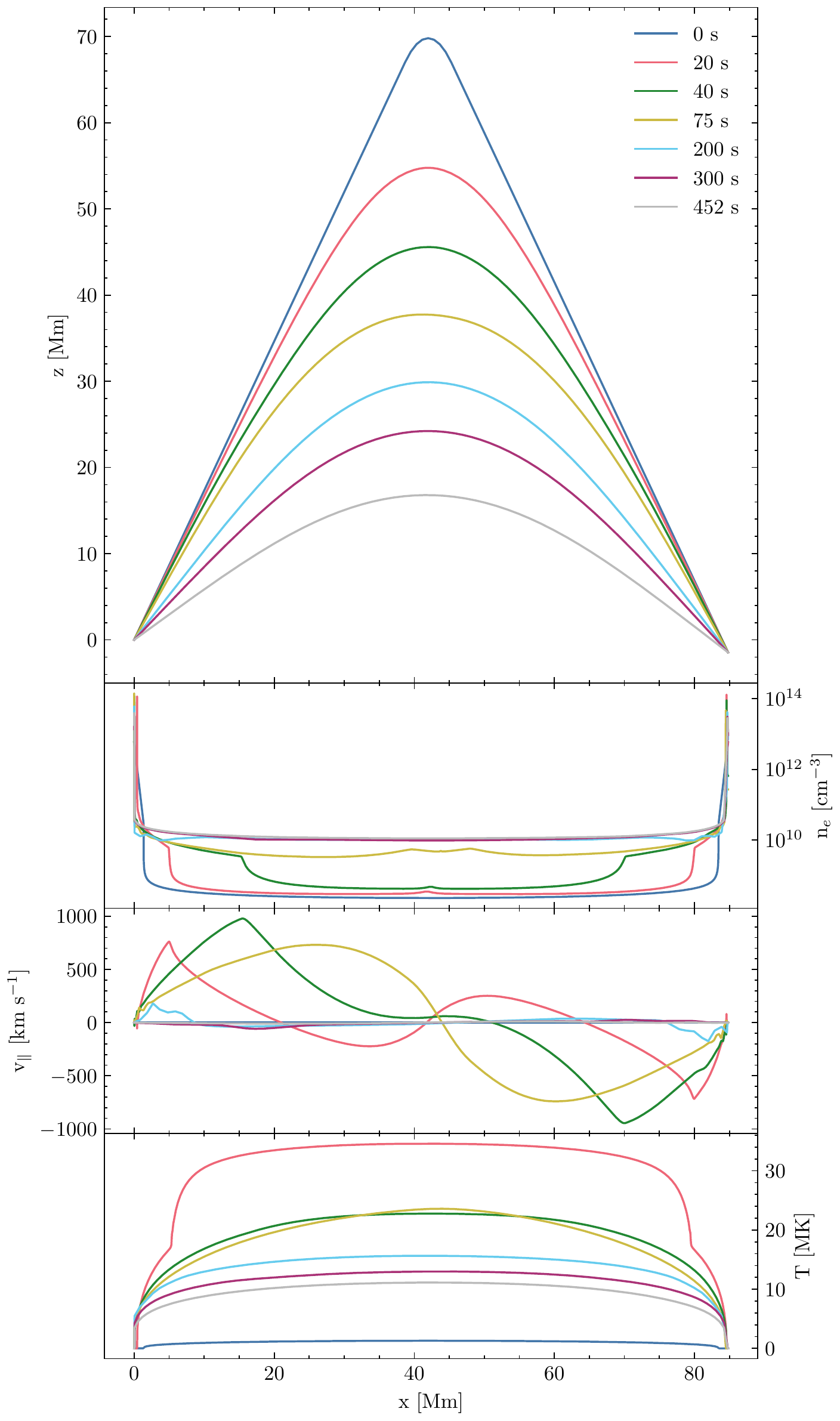}
\caption{Dynamics of the reference simulation over the course of retraction. A face-on view of the tube axis $\textbf{r}(\ell)$ at seven times is shown in the top panel. The following three panels \whaiv{show} the corresponding electron number density $n_e$, velocity parallel to the axis of the tube $v_\parallel$, and temperature $T$. Each panel is plotted against the x-coordinate, scaled such that x=0\,Mm corresponds to the \whaiv{leftmost} boundary of the tube. 
\label{fig:preftraw}}
\end{figure}

The evolution of the tube during its retraction is illustrated in Figure \ref{fig:preftraw}. Starting with an apex at height $z_0$=71\,Mm, the tube retracts until its total length has decreased to $L_f=  93$\,Mm at time $t=452$\,s (7.5 min), yielding a final apex height of $z_f=18$\,Mm. The change in apex height, $\Delta z = z_0-z_f$ = 53\,Mm, is analogous to a post-flare loop moving through a current sheet of the same length \citep[for example]{longcope2018}. Once the retraction is complete, the tube is artificially straightened to lie on the $\hat{\textbf{x}}$-axis, mimicking the effect of the tube reaching the post-flare arcade.  Velocity flows after this point are set to equal the parallel velocity $v_\parallel$ immediately prior to the straightening, and the perpendicular flow is set to zero. The simulation is then run for an additional 32 minutes in the straightened configuration. With no perpendicular flow, the source of \alfven waves is $S_{\pm}=0$, so the wave energies are left to decay freely.

During the retraction, the corona undergoes characteristic flare dynamics. Temperature quickly increases from $T_{\mathrm{co},0}=1.3$\,MK to values well over 30\,MK before decaying down to $\sim 10$\,MK. The parallel velocity along the tube shows that evaporation starts by $t=20$\,s, reaching speeds up to nearly 1\,Mm\,s$^{-1}$. By $t=5$\,minutes, most of $v_\parallel$ has dropped out completely. Once evaporation has begun, the tube's density increases by several orders of magnitude in the corona, and remains elevated after the temperature and velocity in the tube have subsided.

\begin{figure}[t!]
\centering
\includegraphics[width=\columnwidth]{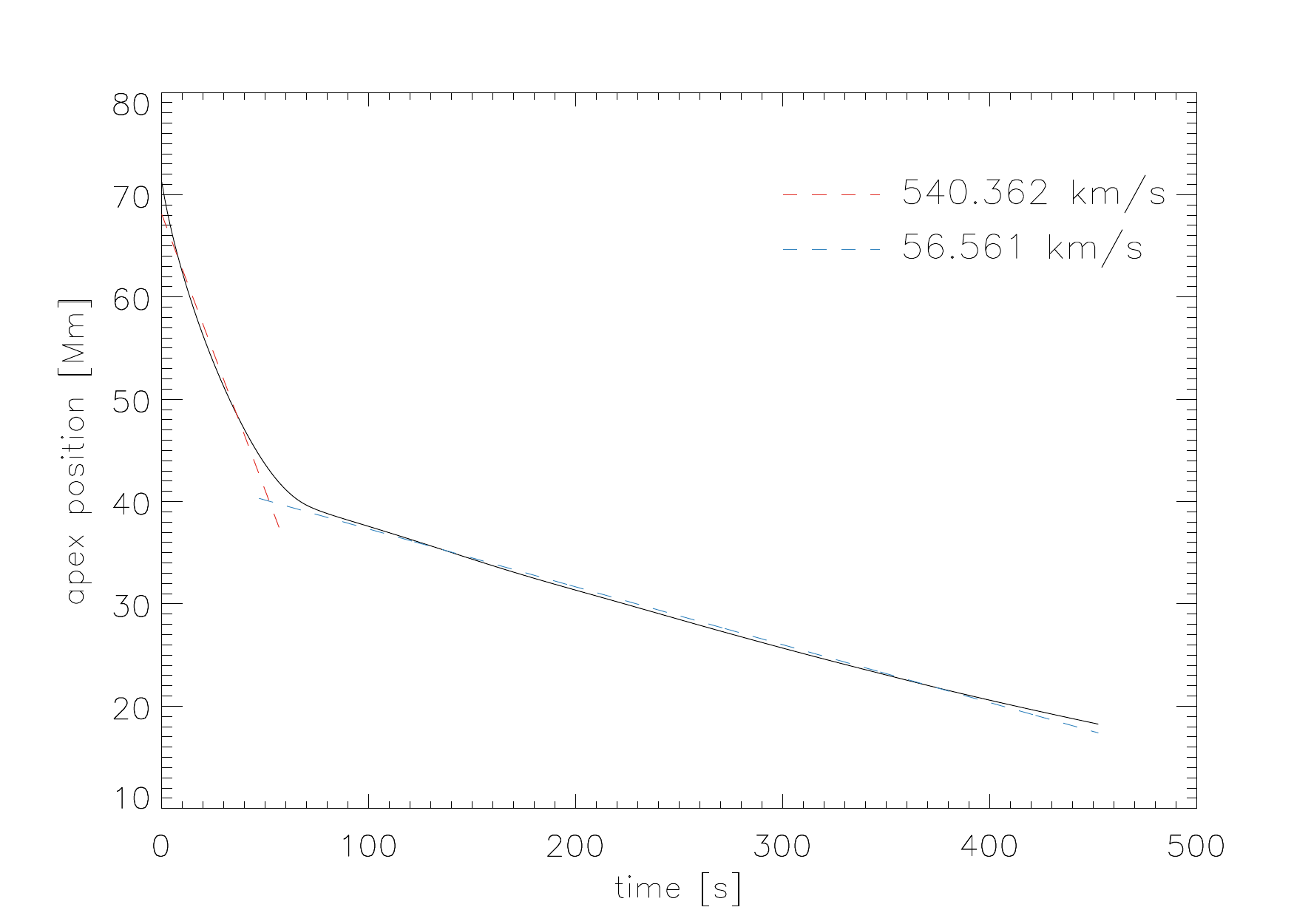}
\caption{Apex position of the flux tube over the retraction period. Linear fits to the fast (red dashed) and slow (blue dashed) retraction phases are over-plotted.
\label{fig:ret_rate}}
\end{figure}

The explosive initial evolution appears more reminiscent of drag-free solutions \citep[e.g.][]{longcope2015}, than the gentle dynamics seen in \cite{unverferth2021}. The effect of drag, however, can be seen from the apparent deceleration of the tube in Figure \ref{fig:preftraw}. To illustrate the change in downward velocity, Figure \ref{fig:ret_rate} plots the apex height of the tube throughout its retraction. Here the retraction appears to occur in two nearly constant-velocity phases; a fast downflow, lasting a little less than one minute, followed by a dramatic reduction of speed that remains constant until the tube is straightened. A straightforward explanation for the deceleration is provided by evaporation. The upward driving of chromospheric material increases the density in the corona, subsequently lowering the local \alfven speed and slowing the retraction.

Linear fits to each phase show the tube initially moving downward at \whaiv{a} speed of 540\,km\,s$^{-1}$ before slowing down to 57\,km\,s$^{-1}$. The initial phase agrees with downflows measured in \cite{longcope2018}, where two-thirds of the 35 downward moving features had velocities lower than 600\,km\,s$^{-1}$. While the features analyzed in their work saw only straight streaks in a height-time map of a current sheet viewed edge-on, observations taken using TRACE\,195\,\AA\ filtergrams saw downflows decelerating in a distinct two-step fashion \citep[for example, see Figure 2 in ][]{sheeley2004}. This deceleration of roughly an order of magnitude occurred prior to the feature reaching the post-flare arcade. Our model thus offers a novel insight into the behavior of these observations.

\begin{figure*}[t!]
\centering
\includegraphics[width=\textwidth]{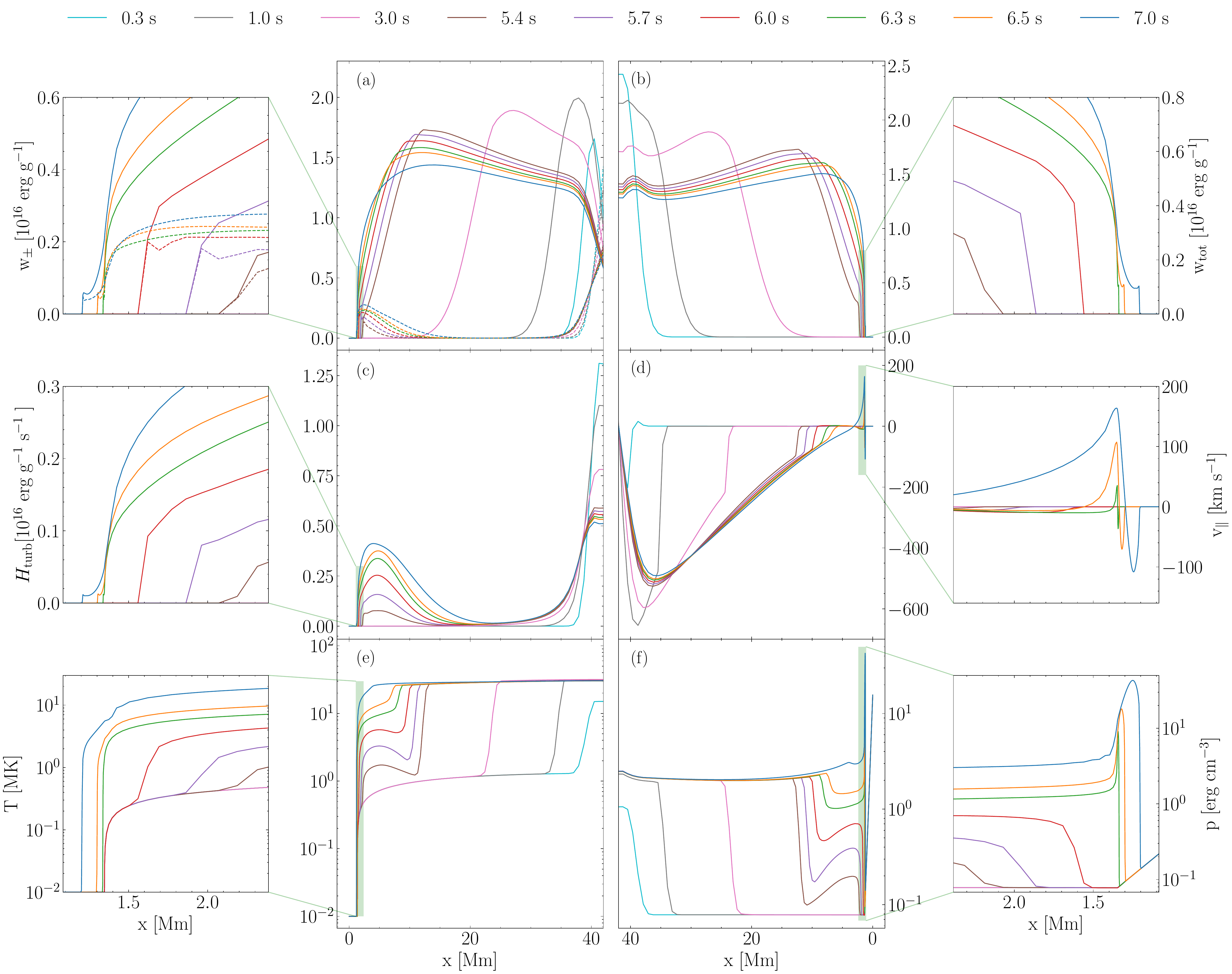}
\caption{Propagation of six flux tube variables over the first seven seconds of the reference simulation. Middle panels show each variable over the entire left half of the loop, while the corresponding leftmost and rightmost panels show enlargements of the chromosphere and upper transition region in detail. Working top to bottom, the variables plotted are (a) the leftward $w^+$ ---solid --- and rightward $w^-$  --- dashed --- propagating \elsasser energies (b) total \elsasser energy $w_\mathrm{tot}$ (c) Heat produced from turbulent energy dissipation $H_\mathrm{turb}$ (d) parallel velocity $v_\parallel$ (e) temperature $T$ and (f) pressure $p$.
\label{fig:tube_prop}}
\end{figure*}

The initial phase of the flux tube's evolution is shown in Figure \ref{fig:tube_prop}, where we see the dynamics of the turbulent \alfven wave energies $w_\pm$. The total energy of the waves $w_\mathrm{tot}$ peaks at $t=0.3$\,s (cyan) and subsequently decreases as the waves propagate towards the footpoints of the loop.  At $t=4.3$\,s, the waves reach the boundary set by the characteristic temperature $T_\mathrm{TR}$=0.5\,MK, corresponding to position $\pm x=2.6$\,Mm from the edge of the tube. Upon reaching the boundary, the waves then begin to reflect, the effects of which can be seen at $t=5.4$\,s.  The solid lines in the upper left panels correspond to the leftward propagating \elsasser energy and the dashed lines correspond to the rightward, counter-propagating energy created from the reflection.  Over times $t>4.3$ s, the point of reflection moves progressively downward as the \whaiv{position where} $T=0.5$\,MK moves by thermal conduction.

Counter-propagating waves appear first at the loop top, where drag creates both waves together.  \whaiv{Interaction between waves} results in significant loop-top heating according to Equation (\ref{eq:hturb}). \whaiv{Turbulent dissipation} drives the temperature of the loop to a peak temperature of $T_\mathrm{peak}=41$\,MK that rapidly drives thermal conduction fronts outward towards the footpoints. The steep temperature gradients shown in the lower left panels of Figure \ref{fig:tube_prop} are evidence of the flux limiter taking effect.

Once the reflection begins ($t>4.3$ s), interactions between \whaiv{the} incident and reflected waves create a second locus of heating near the footpoint.   
A novel result of this simulation is the interaction between the thermal conduction fronts and the increase in temperature driven by the turbulent heating at the reflection boundary. As the conduction fronts move leftward in Figure \ref{fig:tube_prop}, they are impeded by a localized temperature increase in the transition region that grows over time. This growth, initially due to $H_\mathrm{turb}$, is likely compounded by the conduction front, as the initial temperature bump seen at $t=5.4$\,s gradually smooths out with increasing temperature. Nevertheless, the localized heating from reflection creates a separate conduction front\whaiv{, as well as a pressure increase above the transition region.} Remarkably, it is this conduction front that appears to drive evaporation upon reaching the chromosphere, the beginnings of which are seen at $t=6.3$\,s in both parallel velocity and pressure. It should be noted that the leftward movement of the reflection boundary, as shown by the leftward propagation of $w_\pm$, is also due to the localized conduction front, which heats the plasma and moves $T_\mathrm{TR}$ in that direction. 

\begin{figure*}[t!]
\centering
\includegraphics[width=\textwidth]{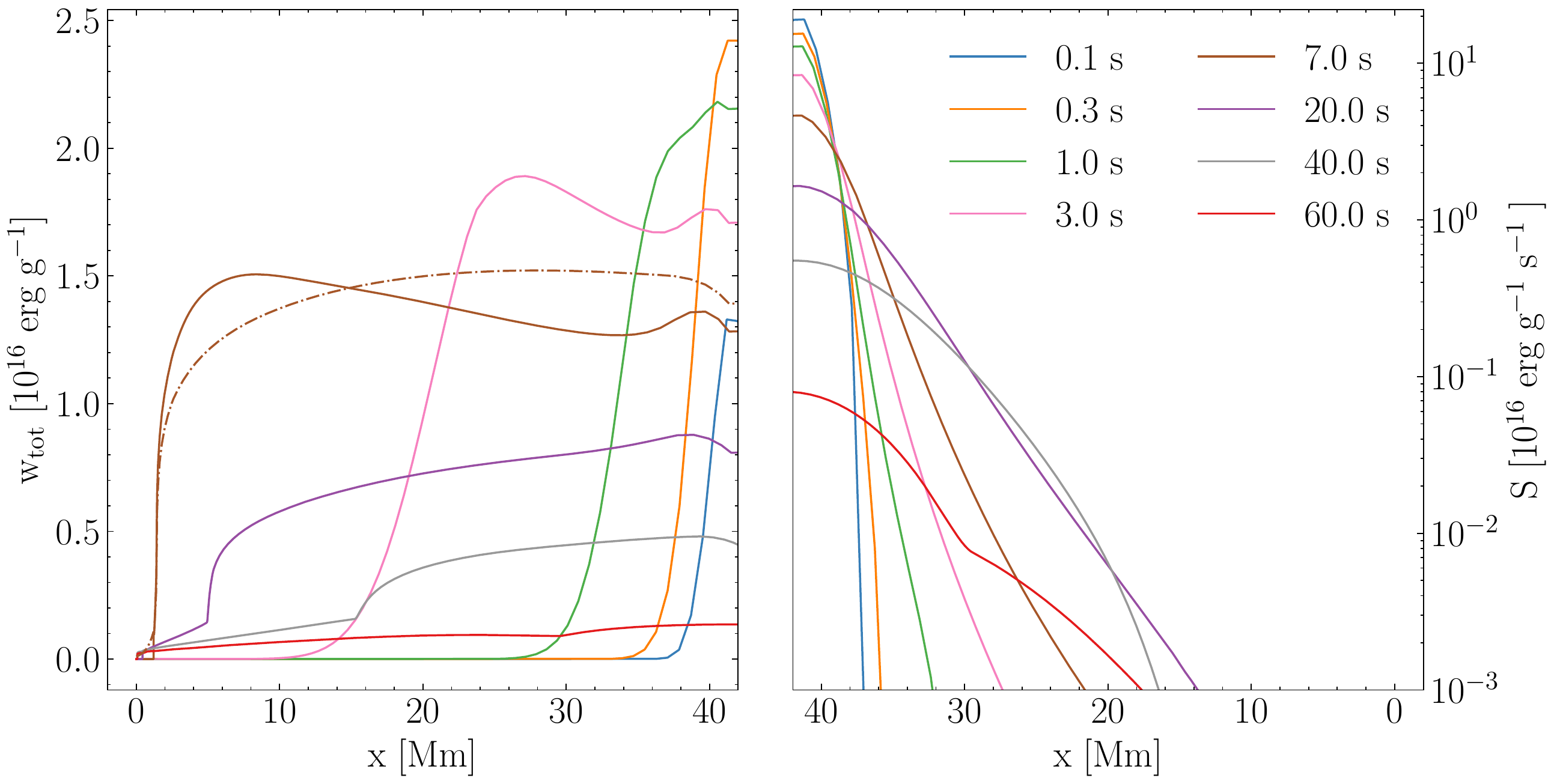}
\caption{Total \elsasser energy $w_\mathrm{tot}$ (left) and the corresponding source from drag power $S$ (right) over the first 60\,s of the reference simulation. An instance of the wave energy envelope per unit mass without sources or reflection (Equation (\ref{eq:env})) is shown as the dashed-dotted curve at $t$=7.0\,s.
\label{fig:w_s}}
\end{figure*}

Figure \ref{fig:w_s} plots the evolution of the total turbulent \alfven wave energies alongside their corresponding sources $S=S_+ + S_-$ for the first minute of the simulation. Here, the amplitude of the source peaks immediately after the retraction begins and is centered around the loop's apex. As the tube continues to retract, the source amplitude decreases by two orders of magnitude and spreads along the extent of the tube. This effect illustrates how the sharp initial bend of the tube decomposes into a more rounded shape via the drag force, subsequently allowing for additional segments of the tube to be deflected downward --- an effect that compounds throughout the retraction. Even so, most of the power in drag remains concentrated at the loop center, leading to a continual creation of turbulent \alfven waves there.

After the \elsasser energies reach their peak, they then decrease along with the source. Of particular note is the shape of the energies themselves, where the amplitude sharply decreases before undergoing reflection at the boundary. To investigate this behavior, we calculate an envelope representing the energy per unit mass in the waves that would otherwise be stationary without sources or reflection. To do this we solve the static version of equation (\ref{eq:w}), with $dw_{\pm}/dt=v_{\parallel}\,\partial w_{\pm}/\partial\ell$, and the left two terms omitted.  The result
\begin{equation}
\label{eq:env}
w_{\pm}^{\rm (env)}(\ell) ~=~ A\, \exp\left[\, \int\limits^{\ell}
\frac{1}{(v_{\rm A}\mp v_{\parallel})}\,
\frac{\partial(v_{\rm A} \pm v_{\parallel}/2)}{\partial\ell'}
\,d\ell'
\, \right],
\end{equation}
is an envelope profile plotted \whaiv{as} a dot-dashed line alongside the normal wave at $t=7$\,s in Figure \ref{fig:w_s}, with arbitrary scaling $A$ adjusted for clarity. The envelope shows a drop in wave energy due mainly to a decrease in the \alfven speed in the transition region. Excess energy density around $x=48$\,Mm is then due to reflection and the production of counter-propagating waves. Similarly, although the source is still relatively high, turbulent dissipation brings the energy density below the envelope for $x\gtrsim$58\,Mm.

\begin{figure}[t!]
\centering
\includegraphics[width=0.66\columnwidth]{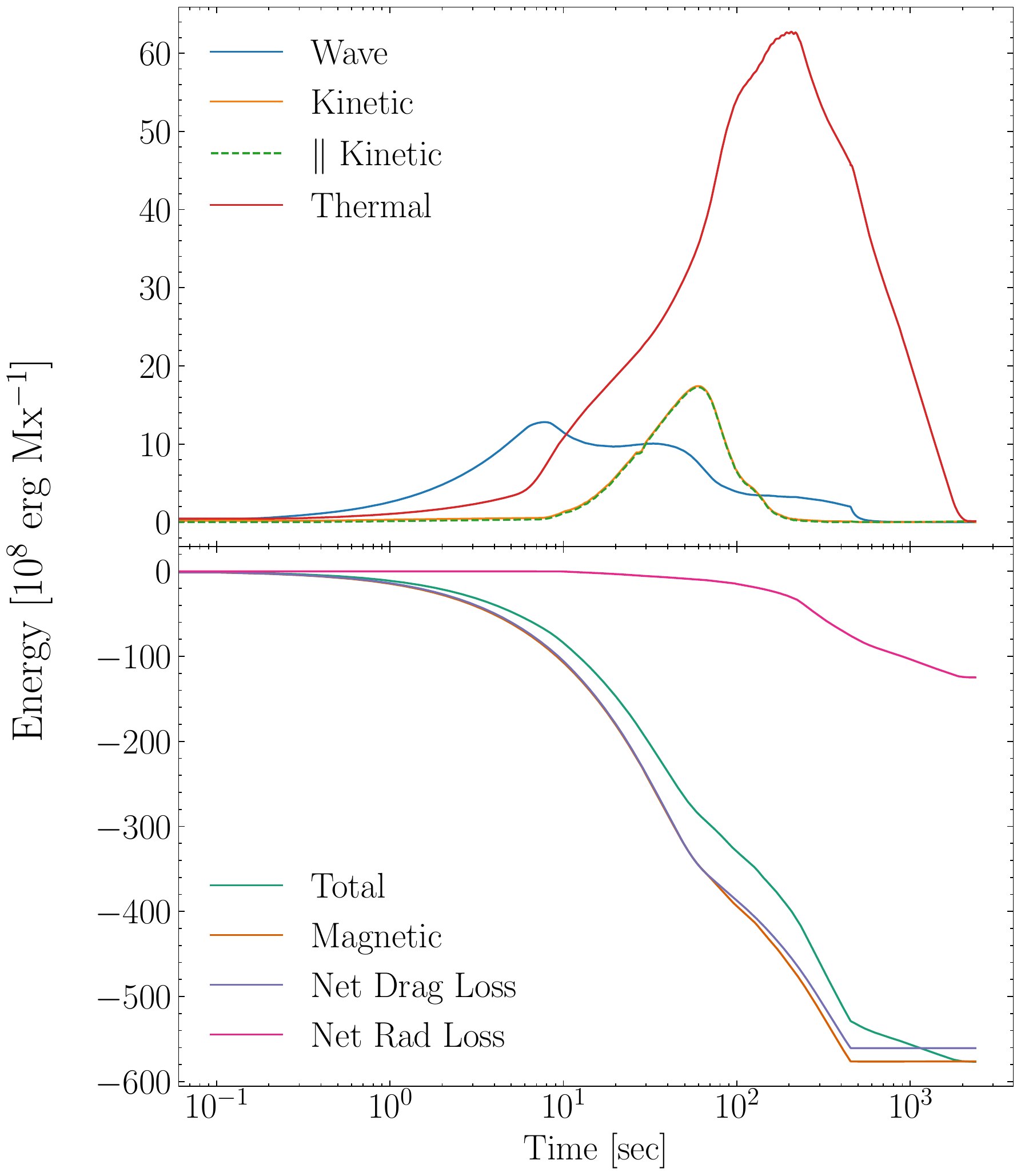}
\caption{Evolution of energies within the tube in units of $10^8$\,erg\,Mx$^{-1}$, plotted against a logarithmic time axis. The bottom panel shows free magnetic energy (dark orange), net drag loss (purple), and radiative losses (magenta). The top panel shows turbulent wave energy (blue), kinetic (orange), parallel kinetic (dashed green), and thermal (red) on an enlarged scale. The total energy (teal) is the sum of the wave, kinetic, thermal, and magnetic energies. \whaiv{Note the kinetic and parallel kinetic energies are virtually indistinguishable.}
\label{fig:wave_erg}}
\end{figure}

The long-term evolution of the tube is shown by the energies plotted in Figure \ref{fig:wave_erg}. The magnetic energy (dark orange) lost from its initial value of $W_{M,0}=1.317\times10^{11}$\,erg\,Mx$^{-1}$ is shown in the bottom panel. Over its retraction, the tube releases a total of $\Delta W_M=5.76\times10^{10}$\,erg\,Mx$^{-1}$ --- 44\% of the initial amount.  Drag (purple) is responsible for removing nearly all free magnetic energy released. The total work done by drag by the end of the retraction is $5.61\times10^{10}$\,erg\,Mx$^{-1}$, constituting over 97\% of the available magnetic energy. The slight deflection seen at $t\sim 60$\,s results from the change in retraction speed, while constant values starting at $t= 452$\,s indicate the tube's straightening. 

The top panel of Figure \ref{fig:wave_erg} shows the remaining energies present in the tube. The total energy found in turbulent \alfven waves (blue) mirrors the drag power, reaching a maximum of $12.8\times10^8$\,erg\,Mx$^{-1}$ before non-linear interactions between counter-propagating waves dominate. The inflection seen in the wave energy at $t=$7.8\,s therefore corresponds to an uptick in thermal energy, shown in red, as the heating from turbulent dissipation increases.  The source of wave energy vanishes once the retraction ends ($t=452$ s), and the wave energy decays rapidly thereafter.  It seems that the non-linear dissipation term works on relatively short time scales provided there is enough reflected power to facilitate non-linear interaction.

The kinetic energy (orange) increases notably at $t=10$\,s, corresponding to the onset of evaporation flows. The parallel kinetic energy (green dashed) is effectively identical to the kinetic energy and makes up 99\% of the total. This is a direct effect of the drag limiting the perpendicular velocity; in drag-free solutions, the parallel flow contributes only a small fraction to the total kinetic energy \citep{longcope2015}.  
Furthermore, thermal energy decreases when the kinetic energy has subsided at $t=206$\,s. This moment, also corresponding to an increase in the radiative losses, can be used to indicate the cooling phase of the loop. By the end of the run, $1.25\times 10^{10}$\,erg\,Mx$^{-1}$ has been lost to radiation.

The importance of wave energy in driving the initial flare evolution is readily seen when compared to the parallel kinetic energy. As alluded to in Section \ref{init}, the energy contained in turbulent \alfven waves is found to be comparable to that contained in parallel plasma flows. Moreover, wave energy in the tube increases immediately upon the start of retraction, whereas kinetic energy increases 10\,s later, after much of the initial dynamics have already taken place. The kinetic energy here is due to flows driven by evaporation --- shown above to be a result of heating via turbulent dissipation at the reflection boundary in the transition region --- and not from plasma deflected by rotational discontinuities traveling at supersonic speeds \citep{longcope2009,guidoni2010}. That is to say, the evolution of energies seen here supports the notion of turbulence acting as the primary mode of heating during \whaiv{impulsive} flare energy release.

\begin{figure}[t!]
\centering
\includegraphics[width=0.66\columnwidth]{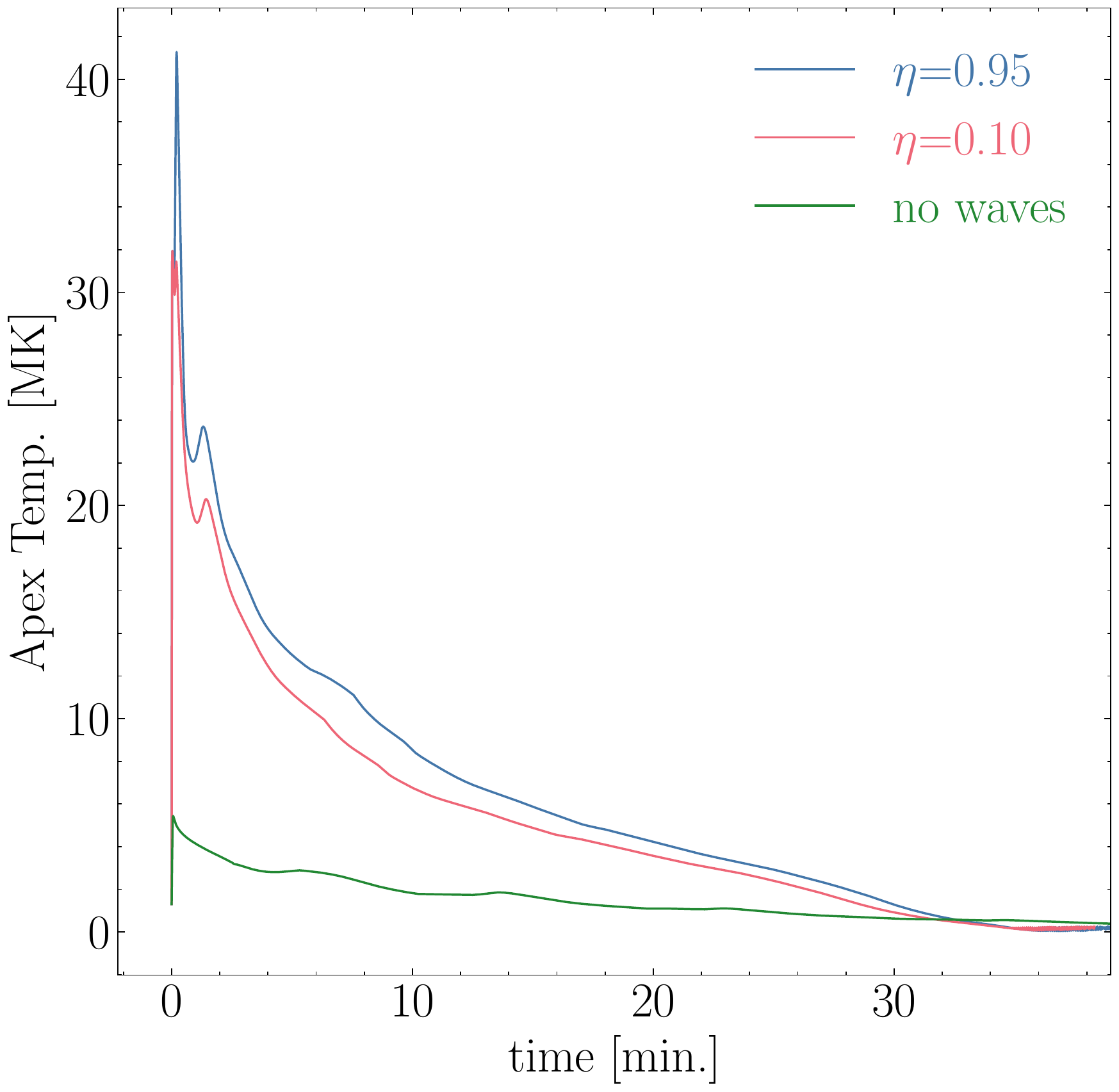}
\caption{Evolution for apex temperature from the results of the reference simulation (blue). Plotted for comparison are apex temperatures for a simulation with a reflection coefficient $\eta=0.10$ (red) and a simulation with no turbulent \alfven waves (green).
\label{fig:tmax}}
\end{figure}

The close relation between explosive flare behavior and turbulent wave energy is also illustrated in the temperature evolution shown in Figure \ref{fig:tmax}. The apex temperature of the reference simulation, shown in blue, is plotted over the entire evolution of the tube. The evolution is characteristic of those observed during flares, reaching a peak of $T=41$\,MK before slowly decaying down to its pre-flare value. In this case, the decay time for the flare is on the order of 35 minutes. We note that the additional peak at $t\sim 2$ minutes is an effect of evaporation. For comparison, a simulation was run using the same drag coefficient, $D$=50\,Mm$^{-1}$, as the reference, but without including turbulent \alfven waves (i.e.\ $f_{\rm turb}=0$). The result, shown in green, only reaches a peak temperature of $T=5.5$\,MK. While the temperature decay is on the same order of time as the reference run, the initial flare evolution is considerably weaker, suggesting kinetic energy alone is insufficient to power strong flare dynamics. 

It seems that the interaction of waves at the loop apex explains much of the peak temperature evolution.  To demonstrate this \whaiv{notion}, we lowered the reflection coefficient to $\eta$=0.10, while leaving the drag and source term the same.  The results, plotted in red on Figure \ref{fig:tmax},  \whaiv{shows} the temperature evolution closely following that of the $\eta$=0.95 run, albeit at slightly lower values, reaching a peak of $T=32$\,MK and decaying to the pre-flare value in virtually the same time. The lower temperatures seen in this case suggest that reflected waves play some role in heating, albeit relatively minor. The \whaiv{small} role played by reflection is consistent with the observation above that non-linear wave dissipation is strong enough to eliminate the wave energy rapidly.  It is evidently also capable \whaiv{of} dissipating the energy over length scales short enough that most does not reach the feet.

\subsection{Effect of Suppressed Thermal Conduction}

We briefly note that the effects of thermal conduction suppression, given by the modification in Equation (\ref{eq:sur}), were found to be negligible. At its largest deviation, conductivity remained within less than 1\% of its classical value.  This is a natural consequence of the low energy density in \alfven waves ($\sqrt{w} \simeq 0.5$ Mm\,s$^{-1}$) relative to the loop's \alfven speed $v_{\rm A}\simeq 10$ Mm\,s$^{-1}$.  The magnetic perturbations will deflect the field lines by a random angle $\sim \sqrt{w}/v_{\rm A}\simeq 0.05$.  This deflection results in only a minor increase in field line \whaiv{path} lengths, resulting in only a minor decrease in diffusivity.

\subsection{Synthetic EUV Emission} \label{syn}

As mentioned in the introduction, a major motivation for including turbulent \alfven waves in the TFT model was to reproduce observations of sustained coronal flare emission.  In order to create a benchmark by which our model can be qualitatively compared to such observations, light curves corresponding to the six coronal EUV channels measured by AIA were synthesized using the results of our reference simulation. 

For a given wavelength channel $i$, the pixel value $p_i$ for the total emission integrated along the flux tube is calculated by
\begin{equation}
    p_i = \int_0^\infty n^2_e(\ell) K_i[T(\ell)]\ d\ell,
\end{equation}
given the temperature-response function $K(T)$ \citep{boerner2012}. These functions were accessed using the \texttt{aia\_get\_response} function in the \textsc{SolarSoft} IDL library \citep{freeland1998}. Because we are only interested in the time evolution, and not the integrated intensity, the light curves are normalized according to their maximum value, making it unnecessary to assign a cross-sectional area to the 1D tube.

\begin{figure*}[t!]
\centering
\includegraphics[width=\textwidth]{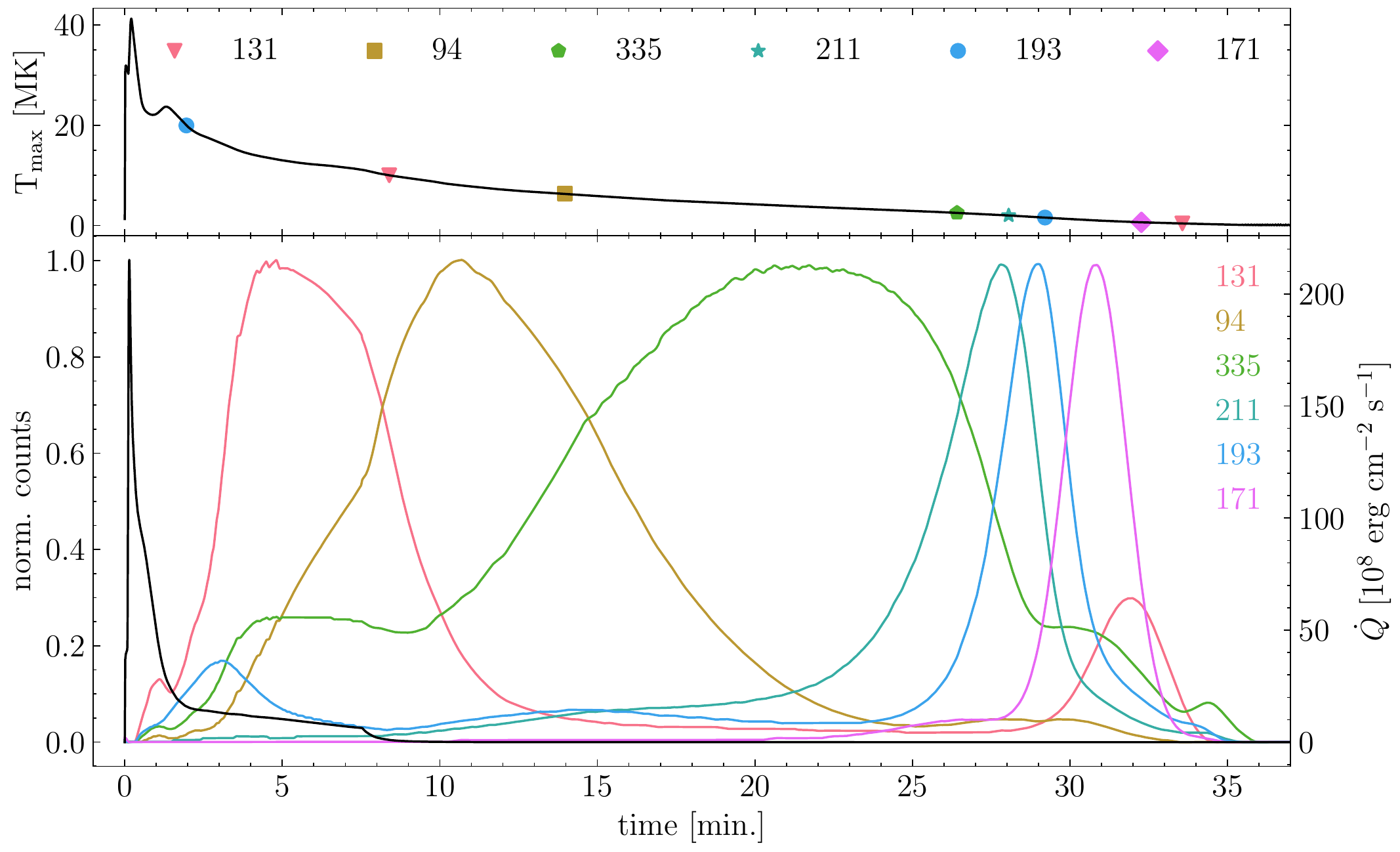}
\caption{Bottom panel: Synthetic light curves of AIA EUV channels 131, 94, 335, 211, 193, and 171\,\AA\ produced using results from the reference simulation. Heating due to the dissipation of turbulent \alfven waves is shown in purple, read against the axis on the right. Top panel: Apex temperature evolution as shown in Figure \ref{fig:tmax}. Symbols denote the times when $T_\mathrm{apex}$ crosses the characteristic formation temperature(s) of the respective AIA channel.
\label{fig:preft2aia}}
\end{figure*}

Synthesized emissions for each AIA EUV channel --- 131, 94, 335, 211, 193, and 171\,\AA --- are shown in Figure \ref{fig:preft2aia}. The lightcurves evolve in a typical fashion and peak successively according to their characteristic temperatures, decaying from 20 to 0.4\,MK. The higher temperature passbands of 131, 94, and 335\,\AA\,, also have noticeably wider emission profiles than the lower temperature passbands of 211, 193, and 171\,\AA. This difference in duration indicates the temperature evolution through these passbands, representing a temperature range of 0.4-2\,MK, occurs more quickly. To be expected, the overall evolution of the light curves agrees well with the temperature decay in Figure \ref{fig:tmax}. If we quantify the decay of the EUV emission by the time it takes the lowest channel in 171\,\AA\ to peak, then the duration of the flare emission in this case is approximately 32 minutes long. 

To illustrate the effect of turbulent \alfven waves on the duration of the EUV emission, the total heat from turbulent dissipation integrated along the tube, $\dot{Q} =  \int H_\mathrm{turb} d\ell$, is plotted alongside the light curves, converted here into a heating rate per unit area. The heating rate roughly tracks the evolution of the total wave energy in Figure \ref{fig:wave_erg}, climbing to a peak of $2.15\times10^{10}$\,erg\,cm$^{-2}$\,s$^{-1}$ in the initial seconds of the simulation before sharply decreasing. Because the loop starts to cool at $t=206$\,s, turbulent heating after this time can be considered the gradual-phase heating supplemental to the impulsive energy release --- analogous to the so-called \textsl{slow-tail} heating used in \citep{qiu2016}. Here, the gradual-phase heating component lasts much less than the duration of the synthesized emission, falling to zero \whaiv{13 minutes after the simulation start time}. Seeking an explanation for extended gradual-phase heating, and longer duration of EUV emission, we revisit the initial parameters of our turbulent transport model in the following section.

\section{Exploring Parameters} \label{sec:ext}

The preceding analysis found the dissipation of turbulent \alfven waves to effectively drive characteristic flare behavior and produce long-duration EUV emissions on the order of 30 minutes. We explore what is required to extend long-term coronal emission by adjusting the parameters of our turbulent transport model. 
Having already visited the reflection coefficient and the modification to the Spitzer-H\"{a}rm conductivity, we now focus on the two remaining parameters: the percentage of drag power converted to turbulent wave energy, $f_\mathrm{turb}$, and the energy correlation length, $\lambda_\perp$.

\subsection{Converted Drag Power}

The amount of energy converted into MHD turbulence from drag losses is dictated by the fraction $f_\mathrm{turb}$ in Equation (\ref{eq:src}). Because the generation of \alfven wave perturbations is not explicitly modeled from the interaction between the flux tube and the current sheet, it is unclear what value of $f_\mathrm{turb}$ would serve as a good approximation for this interaction. We therefore explore the response in turbulent energy to a change in drag loss \whaiv{conversion} by varying the degree of interaction through $f_\mathrm{turb}$. To this end, six simulations are run with $f_\mathrm{turb}$ ranging from 0.05--0.8. The remaining loop parameters match the reference simulation outlined in Section \ref{init}.

\begin{figure*}[t!]
\centering
\includegraphics[width=\textwidth]{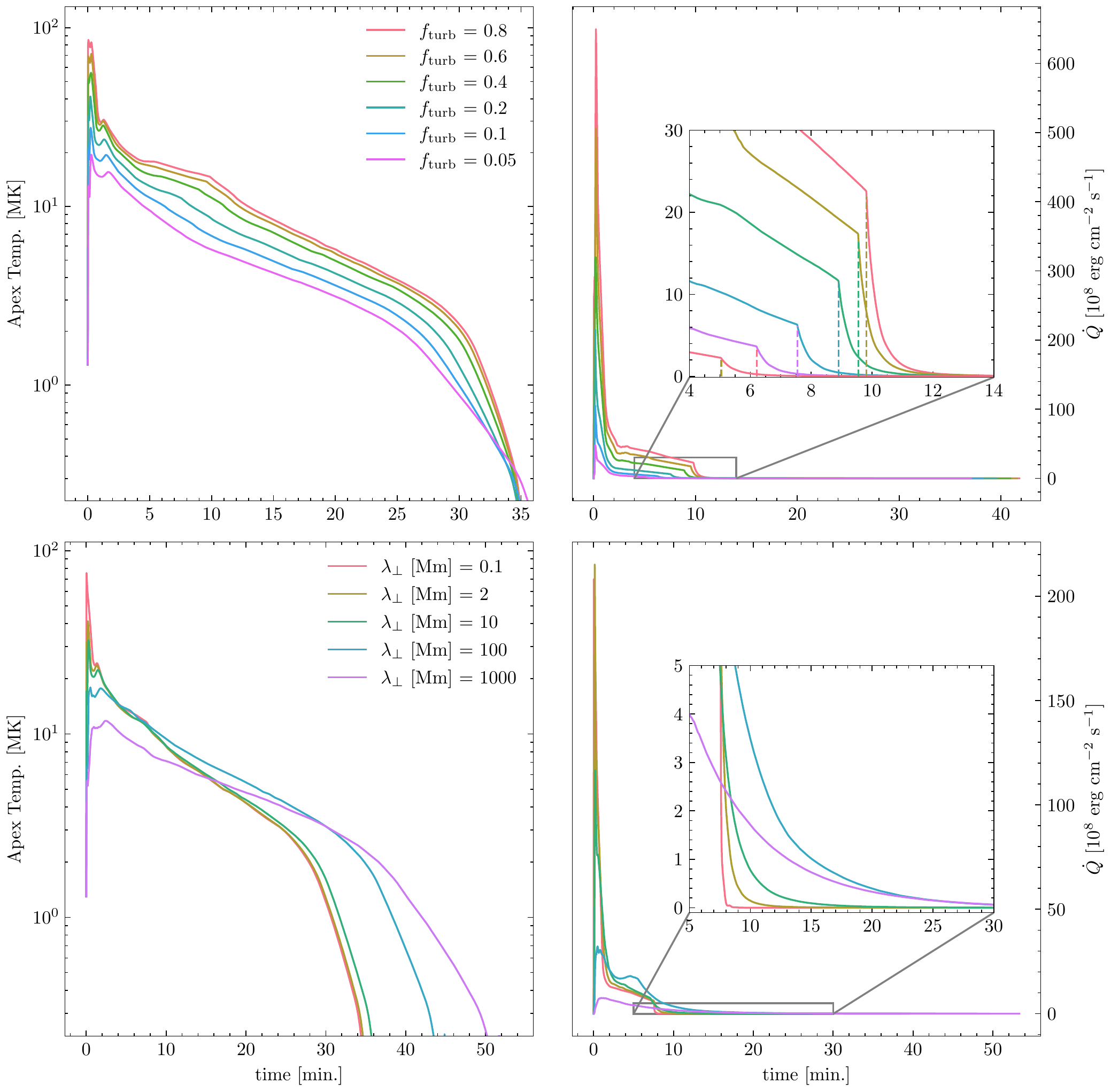}
\caption{Apex temperature evolution (left) and integrated heating rate $\dot{Q}$ (right) for \whaiv{11} simulations run with different values of drag power conversion $f_\mathrm{turb}$ (top) and correlation length $\lambda_\perp$ (bottom). The cases of $f_\mathrm{turb}$=0.2 and $\lambda_\perp$=2\,Mm in the top and bottom panels\whaiv{, respectively,} correspond to the reference simulation analyzed in Section \ref{sec:ref}.  Vertical dashed lines in the upper right panel indicate the time at which retraction ends.
\label{fig:t_turb}}
\end{figure*}

Results from the six simulations are shown in the top two panels of Figure \ref{fig:t_turb}. The apex temperature evolution --- used here as a proxy for the effectiveness of a set of parameters to produce gradual-phase heating --- \whaiv{is} plotted on the left. The overall evolution in each run is similar to that of the reference simulation, shown in green ($f_\mathrm{turb}$=0.2), with temperatures quickly a peak before decaying to ground. In particular, the peak temperature is seen to scale with $f_\mathrm{turb}$. As more drag loss becomes available to be converted into turbulent energy, the tube reaches higher internal temperatures from increased dissipation and subsequent heating. This conclusion is supported by the integrated heating rates $\dot{Q}$, plotted on the right, where higher degrees of drag conversion correspond to larger heating rates. 

Greater degrees of drag conversion were also found to produce more prolonged heating rates. Compared to the $f_\mathrm{turb}$=0.05 case, in which $\dot{Q}$ decayed to zero within 8 minutes, a conversion of $f_\mathrm{turb}$=0.8 resulted in an additional 7 minutes of heating. Regardless, the time taken for the apex temperature to decay to its pre-flare level remained \whaiv{nearly} the same for each of the six simulations, despite increasing heating duration. The similarity between these times is ultimately due to the different evaporation strengths. Stronger evaporation, arising from increased heating following a higher $f_\mathrm{turb}$, leads to a higher density in the coronal segment of the loop, making the losses from radiative cooling more effective and the loop cools faster. As a result, the duration of the temperature evolution appears to be insensitive to the amount of energy put into turbulent \alfven waves.

\subsection{Energy Correlation Length}

The heat generated via turbulence occurs at a rate set by the non-linear interaction between the counter-propagating waves. Because we adopt a one-point closure model, this interaction is characterized by a single parameter, $\lambda_\perp$, indicative of the degree of correlation between the two energy populations. In Section \ref{init}, a range of possible values for $\lambda_\perp$ was motivated from observation. Here, ${\lambda_\perp}$ is taken to be the parameter controlling the efficiency of energy loss from the waves, adjusted from a phenomenological perspective to produce decay timescales that best agree with observation. 

The bottom two panels of Figure \ref{fig:t_turb} show the apex temperature evolution (left) and integrated heating rate (right) for \whaiv{five} simulations run with a range of $\lambda_\perp$=0.1--1000\,Mm. As $\lambda_\perp$ scales inversely with $\dot{Q}$, smaller correlation lengths produce larger heating rates and subsequently larger apex temperatures. Apex temperature evolution is virtually indistinguishable, except during the earliest phase, for all values  $\lambda_\perp \leq 10$\,Mm.  Heating rates for these runs are also similar following the initial phase. 

This similarity appears to be a consequence of energy dissipation within a single transit.  The net energy input into every case is roughly the same.  For cases with $\lambda_\perp \leq 10$\,Mm, that energy is mostly dissipated before reaching the feet, and thus does not remain on the loop beyond the end of retraction at $t=7.5$ min.  This end was noted for the illustration case ($\lambda_{\perp}=2$ Mm), and is repeated in the olive curve.  Those curves with still smaller values of $\lambda_{\perp}$ drop even more rapidly after the termination of retraction (see $\lambda_{\perp}=0.1$ Mm in red).

The two simulations with $\lambda_\perp > 10$\,Mm produced markedly different behavior.  Only in these cases does significant wave energy remain beyond the end of \whaiv{retraction} at $t=7.5$ min (see inset in \whaiv{the} lower right of Figure \ref{fig:t_turb}).  It seems that the non-linear dissipation is small enough for waves to reflect several times.  This persistent wave energy is partly responsible for apex temperatures which decayed to pre-flare levels in 42 and 50 minutes ($\lambda_\perp$ = 100 and 1000\,Mm, respectively) rather than  the 30-minute decay observed in all other cases.  Larger correlation lengths are accompanied by less dissipation in the early phases, resulting in more gentle flare behavior, as evidenced by peak apex temperatures under 20\,MK. The \whaiv{conductive} flux to the feet is therefore smaller, leading to weaker evaporation flows, lower loop density, and  thereby diminished radiative cooling at late times.  This is a second factor extending the cooling time of the loop.  We therefore see that larger correlation lengths, and the weaker non-linear damping they produce,  are critical for our model to extend the duration of coronal emissions.

\subsection{Optimized Simulation}

To capitalize on our better understanding of the role wave energy can play in extending  cooling times we perform one final simulation.  This uses a large correlation length, $\lambda_\perp$=100\,Mm, and high conversion efficiency of $f_\mathrm{turb}$=0.6.  Although the parameter exploration of $f_\mathrm{turb}$ determined that temperature evolution was not sensitive to the degree of drag conversion, a higher value of $f_\mathrm{turb}$ is used to produce higher apex temperatures to agree with flare observations.

\begin{figure*}[t!]
\centering
\includegraphics[width=\textwidth]{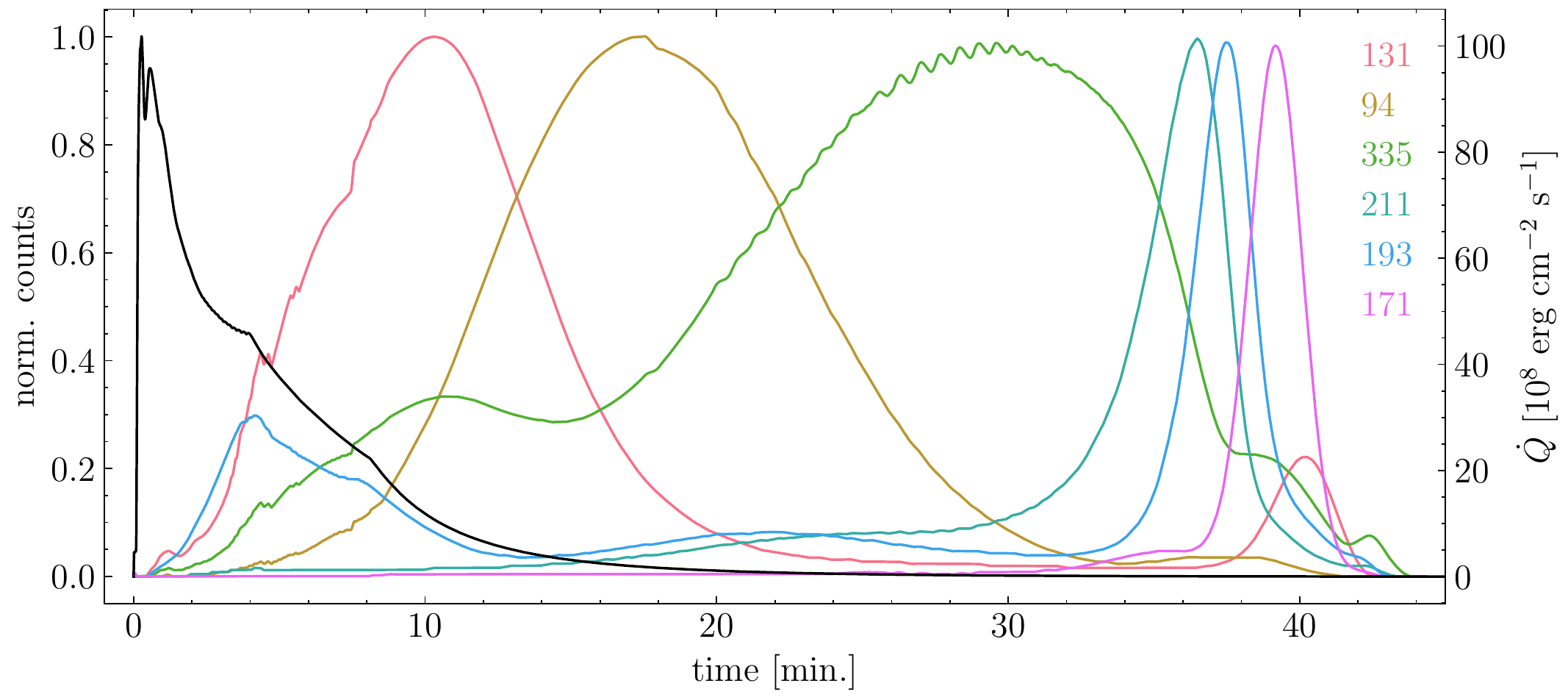}
\caption{Synthetic EUV emission \whaiv{produced from the optimized simulation}, plotted in the same format as Figure \ref{fig:preft2aia}.
\label{fig:s4_preft2aia}}
\end{figure*}

AIA EUV light curves synthesized from this tuned run are shown in Figure \ref{fig:s4_preft2aia}. Compared to the emissions from the reference simulation in Figure \ref{fig:preft2aia}, the profiles here for the hotter channels are broader and peak at later times, illustrating how the temperature remains elevated during this period. The cooler channels, however, are still relatively narrow. At this point in the simulation, the heating rate has decreased from its peak of $1.0 \times 10^{10}$ to $0.2 \times 10^{10}$\,erg\,cm$^{-2}$\,s$^{-1}$, indicating that energy dissipated from turbulence is not enough to counteract energy lost from radiative cooling. 

In this case, the duration of the heating rate lasted 41\,minutes and produced coronal emissions lasting for approximately 40 minutes, given the peak in 171\,\AA. Even though the correlation length and degree of drag loss conversion were 50 and 3 times larger than the reference simulation, respectively, the duration was only 8 minutes longer. \whaiv{Notable differences between the optimized and the reference simulations are summarized in Table \ref{tbl}.} Additional simulations run with increased $f_\mathrm{turb}$ and $\lambda_\perp$, while not shown here, also failed to extend the duration of the emission past 40 minutes, further signifying the balance between strong flare dynamics and gradual turbulent dissipation. As such, the emission duration shown in Figure \ref{fig:s4_preft2aia} represents the upper limit of our turbulent transport model.

\begin{figure*}[t!]
\centering
\includegraphics[width=\textwidth]{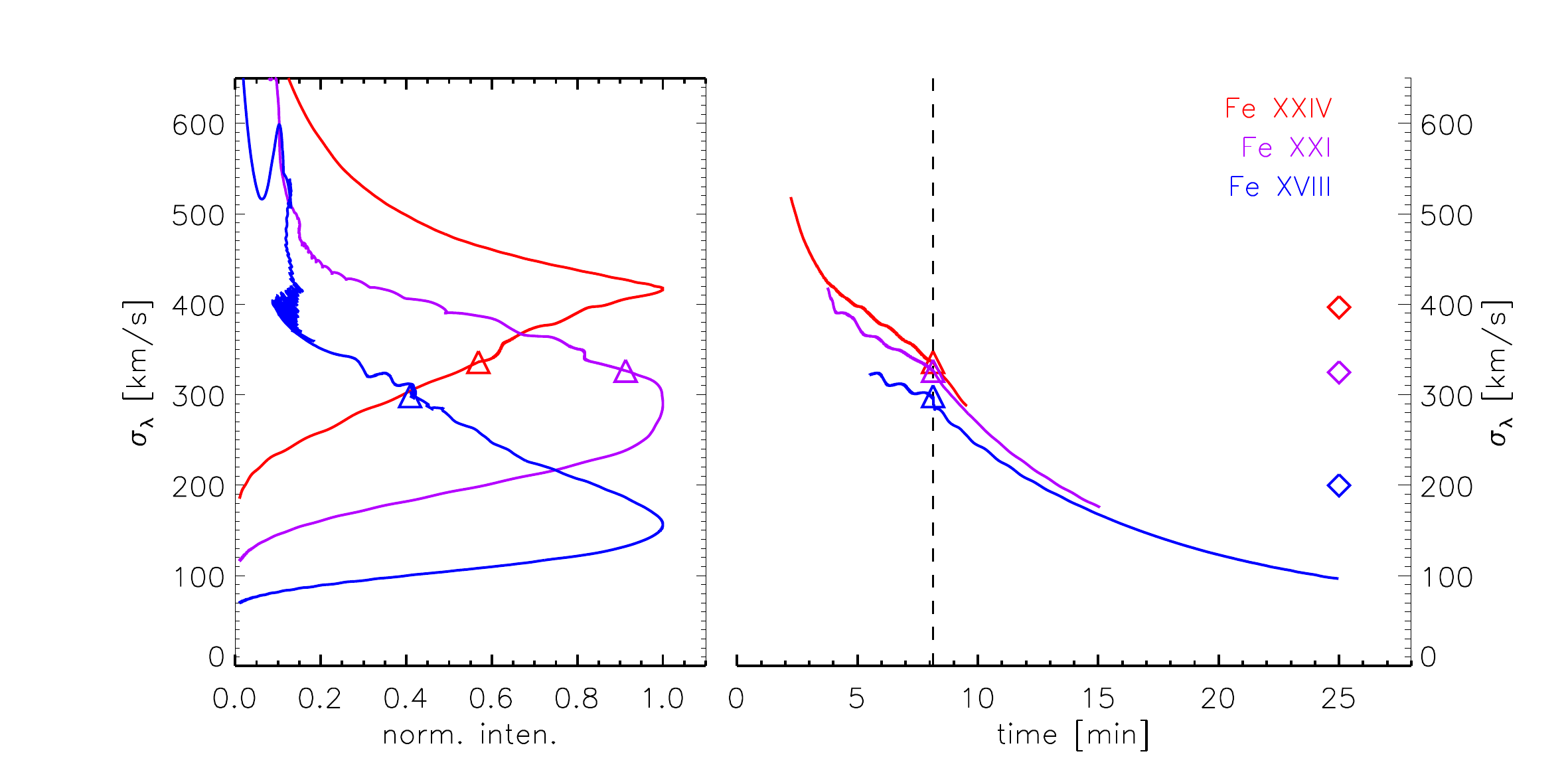}
\caption{Turbulent line broadening, defined in Equation (\ref{eq:sigma}) for spectral lines of Fe {\sc xxiv} 192\AA\ (red),  Fe {\sc xxi} 129 \AA\ (violet), and Fe {\sc xviii} 94 \AA\ (blue).  The left panel plots these against the normalized intensity of the line, and the right against time when the line is bright. \whaiv{Time flows respectively from top to bottom for each curve in the left panel.} The time of loop straightening ($t=8.13$ min) is indicated by a vertical dashed line and triangles on each curve.  Diamonds along the right axis are the average values for each line. 
\label{fig:vnt}}
\end{figure*}

\alfven\ wave turbulence should be evident in \whaiv{an} excess broadening of hot spectral lines.  The turbulent energy per mass gives the turbulent velocity squared averaged over all fluctuation time scales: 
\begin{equation}
    w_{\rm tot} = w_++w_-=\frac{1}{2} \langle |{\bf v}|^2\rangle +
    \frac{1}{2}\left\langle \frac{|{\bf b}|^2}{4\pi}\right\rangle=  \langle |{\bf v}|^2\rangle,
\end{equation}
after assuming turbulence is Alfv\'enic.  The net effect on an unresolved observation of spectral line $\lambda$ is characterized by a spatial average weighted by the local intensity of the line, $I_{\lambda}(\ell)$,
\begin{equation}
    \bar{w}_{\lambda} = \left(\int I_{\lambda}(\ell)\, d\ell \right)^{-1}\,
   \int w_{\rm tot}(\ell)\,I_{\lambda}(\ell)\, d\ell.
\end{equation}
If the axis of the tube is viewed perpendicularly, then the averaged broadening of \whaiv{the} spectral line $\lambda$ along the line of sight, say ${\bf \hat{z}}$, is
\begin{equation} 
\label{eq:sigma}
\sigma_{\lambda}=
\sqrt{\langle v_{z}^2\rangle}=\sqrt{\frac{\bar{w}_\lambda}{2}}.
\end{equation}
Figure \ref{fig:vnt} shows the broadening of three lines typically associated with flares, both versus time (right) and normalized line intensity (left).  Cooler lines show lower broadening, and all decrease with time as the turbulence decays.

\section{Discussion} \label{sec:dis}

We have extended a post-reconnection flare model to include the production and  dissipation of turbulent \alfven waves. 
This extension involved a system of turbulent transport equations inspired by existing models of solar wind turbulence. One-dimensional simulations tracked the evolution of MHD turbulence through their aggregate energy densities, which remained trapped in the corona by reflection from high-density gradients in the transition region. Turbulent energy dissipated through the non-linear interaction between counter-propagating waves and produced heating longer than previous versions \whaiv{of the \textsc{PREFT} flare code}. Synthetic light curves from AIA EUV bands showed increased similarity with typical flare observations.

Ours is the first model of its kind to generate \alfven\ wave turbulence from reconnection outflow and model its effect on long-term flare signatures.  In the spirit of a first attempt, we chose to represent the physical processes involved using simple, fixed parameters.  Energy is removed from the retracting flux by a simple aerodynamic drag force, parameterized by the parameter $D$.  We chose the value $D=50\,{\rm Mm}^{-1}$ in order to slow the retraction to a level consistent with typical observations.  A fraction, $f_{\rm turb}$, of the energy lost to drag appears as unresolved Alfv\'en wave turbulence.  While this process is likely to be very complicated, with a \whaiv{fraction} varying in time as well as space, we chose to use a fixed value of $f_{\rm turb}$.  We performed a series of runs and different values of $f_{\rm turb}$ and found \whaiv{its} most important effect was in setting the peak flare temperature.  

The physics of \alfven\ wave propagation, reflection, and dissipation is also captured through a small number of free parameters.  Reflection is modeled through a reflection coefficient, $\eta$, applied at the point where $T=T_{\rm TR}=0.5$ MK.  Wave dissipation occurs through non-linear dissipation involving waves of both species.  This process is parameterized through an effective correlation length, $\lambda_{\perp}$, which was varied to explore \whaiv{its} effects on the flare evolution.

The duration of turbulent heating was found to be most dependent on the correlation length $\lambda_\perp$. 
Correlation lengths $\lambda_\perp \leq 10$\,Mm lead to very rapid dissipation and wave energy vanishes rapidly after retraction ends.  Only in those with $\lambda_\perp \gtrsim 100$\,Mm does wave energy persist beyond the end of retraction, thereby prolonging the cooling of the flare.  If \alfven\ waves are actually responsible for the long flare cooling times observed, non-linear wave dissipation would need to be characterized by such large correlation lengths.

The finding described above requires reconsideration of the physical significance behind the parameter $\lambda_{\perp}$.  Our turbulence model assumes the small-scale fluctuations responsible for the MHD turbulence to be on scales smaller than the radius of the flux tube, so $\lambda_{\perp}$ cannot be simply taken as the correlation length of the turbulence.  It was mentioned above that our parameter incorporates the correlation between the counter-propagating waves, such that the true correlation length is 
$\lambda'_{\perp}=\lambda_{\perp}/\xi$, where $\xi$ is the fraction of the counter-propagating turbulence which is actually uncorrelated. For an effective correlation length of $\lambda_\perp=100$\,Mm and a tube with radius $R=\lambda_{\perp}=$2\,Mm, this gives $\xi$=0.02, meaning that only 2\% of the counter-propagating waves are uncorrelated and able to produce a turbulent cascade. While an imperfect correlation should be expected between the two energy populations, it is worth noting that the length scale for the small-scale fluctuations (e.g. $R$) and the length scale for the energy correlation do not necessarily need to be related \citep{zank2012}. Should this perspective be taken into account, it would also substantiate the range of length scales derived in Section \ref{init} using the decay rates of non-thermal broadenings observed in  Fe\,\textsc{xxiv}\,\AA\ spectral lines, where $\lambda_\perp \sim $30-150\,Mm.

The heating rates produced in this work also fit the description of a two-step heating profile as described in \cite{qiu2016}. The initial phase of turbulent dissipation was highly impulsive in all cases, reaching a peak within 20\,s before quickly decaying. Following the impulsive energy release, defined here by the time it takes a loop to enter a regime of increased radiative losses, heating from dissipation then became more gradual and lasted anywhere from 5 to 50 minutes. In particular, the heating rate produced in Figure \ref{fig:s4_preft2aia} was found to match the general heating profile described in \cite{zhu2018}, where energy deposited during the impulsive and gradual phases was required to be 60\% and 40\% of the total energy released in order to reproduce sustained EUV emission. Here, after an impulsive phase that lasted 240\,s, the gradual phase released 46\% of the total turbulent wave energy over 36 minutes. To the best of our knowledge, this mechanism was the first to reproduce these two-step heating rates in a self-consistent manner.

A byproduct of using turbulent \alfven waves for gradual-phase heating was their capacity to be equally effective at driving characteristic flare behavior during the initial phase of the simulation. By inducing turbulence from drag, we were able to reintroduce energy back into the tube that would have otherwise been lost to the background of our flux tube. Hence, our model successfully reproduced impulsive flare behavior while still keeping the speed of reconnection outflows within their observed ranges. Investigations into the earlier phases of turbulent dynamics presented here, such as the forward-modeling of Fe\,\textsc{xxi}\,\AA\ lines, may explain the large broadenings of such lines seen during flares \citep{polito2019} and could serve to constrain our model.

While our model produced long-duration heating in conjunction with impulsive flare dynamics, it was not able to reproduce sustained EUV emission on the order of hours, as seen in observation. One possible explanation for this discrepancy is the value of the drag coefficient $D$=50\,Mm$^{-1}$ used in this paper. Increasing $D$ would lead to lower retraction rates, likely resulting in more prolonged heating and more energy available to convert to turbulent \alfven waves. Because the first stage of our reconnection downflow was observed to be 540\,km\,s$^{-1}$, the drag coefficient could be increased to the point of generating retraction rates on the order of 50\,km\,s$^{-1}$ --- the lower speed limit of SADs observed in \cite{savage2012}. 

In this new model, the chromospheric response to reconnection is driven in two different ways.  The first is thermal conduction, which was part of even the earliest flare loop models \citep{nagai1980,pallavicini1983}.  Augmenting this, \alfven\ waves are created by retraction at the loop apex and propagate to the chromosphere where they are dissipated by interacting with their own reflection.  This second transport channel drives chromospheric evaporation beyond that attributable to conduction alone.  It is worth noting that a fraction $1-f_{\rm turb}$ of the drag work will appear as turbulence on field lines neighboring that just reconnected.  Some of the neighboring flux will still be unreconnected, and will now experience some chromospheric evaporation {\em before} reconnecting.  There are several lines of observational evidence suggesting that flare reconnection occurs on field lines whose density is higher than one would expect in the ambient, pre-flare corona \citep{veronig2004,veronig2005,guo2012}.  Our model offers a possible explanation for this previously puzzling fact.

Another mechanism proposed to extend the  cooling time of a flare loop is through suppression of thermal conduction.  Previous investigations into the evolution of flare loops using zero-dimensional models have shown that alterations to the mean-free path of thermal electrons extended cooling times considerably \citep{zhu2018,bian2018}.  Our model accounts, in Equation (\ref{eq:sur}), for a single MHD mechanism for \whaiv{decreasing} $\kappa$: decreasing the mean axial step size $\ell_z$ by the lengthening of the actual path, $\ell$, due to field line meandering on sub-resolution scales, i.e.\ MHD turbulence.  When the perturbation velocity is significantly lower than the local \alfven\ speed, as implied by observations and explicitly found in our model, then field lines are lengthened only slightly.  The result is the very small reduction in $\kappa$ we have observed.  Effective reduction in the mean-free path must occur not through MHD mechanisms but through the kinds of particle kinetics effects suggested by others \citep{Bian2016}.

The level of turbulence predicted here can be compared to observation through turbulent line broadening of high-temperature spectral lines.  The level in our optimized run produces $\sigma$ between 100--400 km/s, a factor of 2-3 higher than typically observed \citep{warren2018,tian2014,polito2019}.  There are, however, many reasons to expect our estimate to be too large.  The estimate in Equation (\ref{eq:sigma}) assumes the field lines \whaiv{are} viewed exactly perpendicularly and that the observation integrates over all turbulent frequencies.  The turbulence consists of \alfven\ waves on the loop, whose frequencies will extend down to the fundamental with a period of two \alfven\ transit times --- perhaps up to 20 seconds.  Integration times shorter than this will omit the lower frequency component of the spectrum, which contains the most energy.  This will reduce the value of $\sigma$ actually observed.  It should be noted that the turbulent broadening in our model naturally decreases on an approximately 15-minute scale.  This is consistent with observations \citep{kontar2017,warren2018}.

The model presented here is self-consistent to the extent that turbulence is ultimately generated by the retraction of a flux tube through a current sheet. The interaction between the current sheet and the tube responsible for both the drag force and the excitement of turbulent \alfven waves, however, remains an assumption of our model and is not directly investigated here. Moreover, the resolution of \textsc{PREFT} required MHD turbulence to be modeled according to its aggregate energy density. Small-scale perturbations in the velocity and magnetic fields that constitute the turbulence were not considered. While the simplicity of the turbulent transport model in this work warrants further study, any improvements would likely require techniques beyond the formalism of 1D flare loop simulations. Despite these limitations, the results presented here illustrate the efficacy of MHD turbulence as a mechanism for flare energy transport and heating, and provides an additional tool by which other flare phenomena can be studied.




\bibliography{short_abbrevs,ref}{}
\bibliographystyle{aasjournal}

\end{document}